\documentclass[onecolumn,11pt]{IEEEtran}
\usepackage{times}
\usepackage{cite}
\usepackage{color}
\usepackage{epsf}
\usepackage{epsfig}
\usepackage{graphicx}
\usepackage{graphics}
\usepackage{epstopdf}
\usepackage[small]{caption}
\usepackage{amsmath}
\usepackage{amssymb}
\usepackage{amsthm}
\usepackage{amsxtra}
\usepackage[ruled]{algorithm2e}
\usepackage{algorithmic}
\usepackage{enumerate}
\usepackage{multirow}
\usepackage{enumerate}
\usepackage{amssymb}
\usepackage{lipsum}
\usepackage{setspace}
\usepackage{multirow}
\usepackage{subcaption}
\usepackage{url}

\usepackage{rotating} 
\usepackage{graphicx} 

\newcommand{\Rmnum}[1]{\expandafter\@slowromancap\romannumeral #1@}

\makeatother
\begin{document}
%
\title{Transforming Energy Networks via Peer to Peer Energy Trading: Potential of Game Theoretic Approaches}
\author{Wayes~Tushar,~\IEEEmembership{Senior Member,~IEEE,}~Chau~Yuen,~\IEEEmembership{Senior Member,~IEEE,}~Hamed Mohsenian-Rad,~\IEEEmembership{Senior Member,~IEEE,}~Tapan Saha,~\IEEEmembership{Senior Member,~IEEE,}~H. Vincent Poor,~\IEEEmembership{Fellow,~IEEE} and~Kristin L. Wood
\thanks{W. Tushar and T. Saha are with the School of Information Technology and Electrical Engineering of the University of Queensland, QLD, Australia. (Email: w.tushar@uq.edu.au; saha@itee.uq.edu.au).}
\thanks{C. Yuen, and K. L. Wood are with the Singapore University of Technology and Design (SUTD), 8 Somapah Road, Singapore 487372. (Email:\{yuenchau, kristinwood\}@sutd.edu.sg).}
\thanks{H. Mohsenian-Rad is with the Department of Electrical Engineering of the University of California at Riverside, CA, USA. (Email: hamed@ee.ucr.edu).}
\thanks{H. V. Poor is with the Department of Electrical Engineering of Princeton University, Princeton, NJ, USA. (Email: poor@princeton.edu).}
}\IEEEoverridecommandlockouts
\maketitle\doublespace
\begin{abstract}
Peer-to-peer (P2P) energy trading has emerged as a next-generation energy management mechanism for the smart grid that enables each prosumer of the network to participate in energy trading with one another and the grid. This poses a significant challenge in terms of modeling the decision-making process of each participant with conflicting interest and motivating prosumers to participate in energy trading and to cooperate, if necessary, for achieving different energy management goals. Therefore, such decision-making process needs to be built on solid mathematical and signal processing tools that can ensure an efficient operation of the smart grid. This paper provides an overview of the use of game theoretic approaches for P2P energy trading as a feasible and effective means of energy management. As such, we discuss various games and auction theoretic approaches by following a systematic classification to provide information on the importance of game theory for smart energy research. Then, the paper focuses on the P2P energy trading describing its key features and giving an introduction to an existing P2P testbed. Further, the paper zooms into the detail of some specific game and auction theoretic models that have recently been used in P2P energy trading and discusses some important finding of these schemes.\end{abstract}
 \setcounter{page}{1}
\section{Background and Motivation}\label{sec:background}
Due to growing concerns for environmental sustainability and climate change, there has been a constant pursuit for an alternative energy system, in which energy production, transmission, distribution, and consumption would take place in an environmentally sustainable fashion. As a result, development of smart, sustainable, and green solutions including widespread deployment of distributed energy resources (DERs) in residential houses~\cite{Tushar-TIE:2014}, introduction of electric vehicles (EV) on roads~\cite{Gan_TPS_2013}, and the establishment of various smart energy services such as demand response management~\cite{Mohsenian-Rad_TSG_2010} for effectively managing energy within the electricity grid are being given preeminent importance recently. Consequently, different signal processing techniques have been used in the last decade to bring these solutions to the forefront of the consumers. Examples of such signal processing techniques include, but not limited to, machine learning, artificial intelligence~\cite{Boss_IEEE_2017}, and game theory~\cite{Saad_GameSmartGrid_2012}. 

An important objective of using these signal processing techniques is promoting the use of renewable energy sources within the energy grid. For example, machine learning and artificial intelligence have been extensively used to forecast the power generation from solar panels and wind turbines~\cite{Voyant_RE_2017}. Due to such innovative use of signal processing tools, and extensive rebates from local governments, a number of techniques are established at this moment to utilize the DERs as the main or subsidiary source of energy across the globe. In particular, the global market for rooftop solar panels is booming. For instance, whereas the global market for rooftop solar panels was nearly US$\$ 30$ billion in $2016$, it is expected to grow by $11$ percent over the next six years~\cite{Peck_IEEE_Spectrum_2017}. Meanwhile, the shift towards solar is being complemented by an increase in the adoption of residential energy storage systems, whose ability to deliver is predicted to grow from around $95$ megawatts (MW) in $2016$ to more than $3,700$ MW by $2025$~\cite{Peck_IEEE_Spectrum_2017}.
\begin{figure*}[t]
\centering
\captionsetup{justification=centering}
\begin{minipage}[l]{\textwidth}
\centering
\includegraphics[width=0.6\columnwidth]{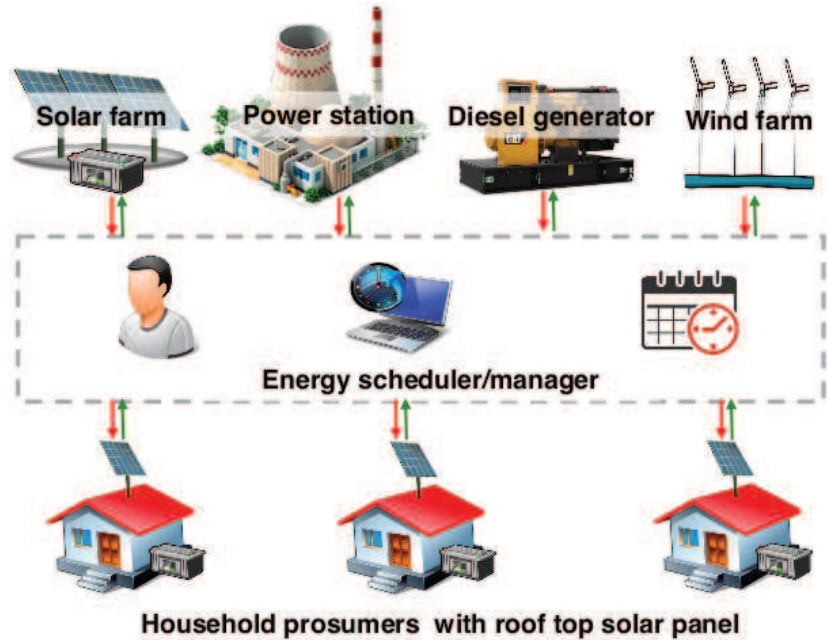}
\subcaption{\footnotesize{Demonstration of traditional FiT scheme.}}
\label{fig:FeedInTariff}
\end{minipage}
\\
\begin{minipage}[r]{\textwidth}
\centering
\captionsetup{justification=centering}
\includegraphics[width=0.9\columnwidth]{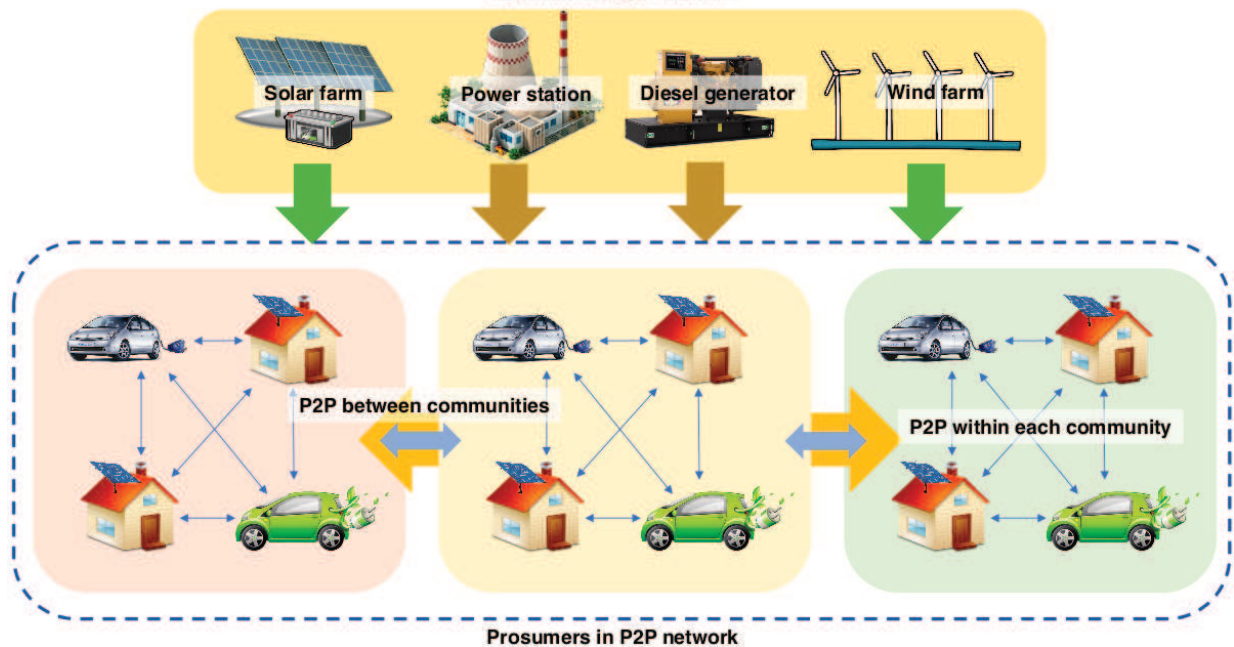}
\subcaption{\footnotesize{Demonstration of how a P2P energy trading network may contribute to alleviate dependency on the main grid.}}
\label{fig:figure_P2P}
\end{minipage}
\captionsetup{justification=centering}
\caption{This figure provide an overview of current trend of renewable energy trading in the smart grid and emerging P2P energy trading scheme.}
\label{fig:CurrentTrendTrading}
\end{figure*}

Hence, if properly utilized, these energy sources at the edge of the grid could help manage demand more efficiently. Nonetheless, this will only happen if the people own these generating assets are fully incorporated into the energy market~\cite{Peck_IEEE_Spectrum_2017}. To this end, feed-in-Tariff (FiT) scheme is a suitable example that engages customers to participate in the energy trading in the market. In FiT, as shown in Fig.~\ref{fig:FeedInTariff}, prosumers\footnote{Prosumers are the energy consumers who also produce electricity.} with DERs such as roof-top solar panel sell their excess solar energy only to the grid, and can also buy energy from the grid in case of any energy deficiency. However, due to the significant difference between the buying and selling prices per unit of energy, the benefit to the prosumers from participating such trading of energy is not significant. As a result, some of the FiT techniques are now discontinued~\cite{FiTPolicy_2015}. Under such circumstances, the creation of new energy market that would allow small-scale participants (or user) actively trade energy with one another in real time, and thus facilitate a sustainable and reliable balance between generation and consumption of energy within the community has become significantly important~\cite{Mengelkamp_AE_2017}.

As such, peer-to-peer (P2P) energy trading is being considered as a potential tool to promote the use of DERs within the energy grid~\cite{Mengelkamp_AE_2017}. The main objective of P2P sharing is to break the centralized infrastructure of the electricity grid by allowing the direct communication and supply of energy between various prosumers with DERs within the energy system, as shown in Fig.~\ref{fig:figure_P2P}. This enables the interested consumers to buy renewable energy at a cheaper rate from a peer (or neighbor) with excess renewable energy (e.g., from rooftop solar), and thus reduce its dependency on the grid or central supplier~\cite{Economist_2013}. Development of such P2P energy trading has significant potential to benefit the prosumers in terms of both earning revenues and reducing electricity cost, as well as lowering its dependency on the grid. An example of recent development of such P2P technology in real energy system can be found in the Brooklyn Microgrid~\cite{Mengelkamp_AE_2017}.

Direct involvement of each user in energy trading with one another and with the grid makes P2P system uniquely different from existing FiT. It poses the challenge of modeling the decision-making process of each participant for the greater benefit of the entire energy network while taking human factors such as rationality, motivation, and environmental friendliness into account. Particularly, in settings where there are many users with conflicting interests participate, it would be quite challenging either to capture such conflicting interests in designing the decision-making process of each participant, or to motivate them to cooperate, if necessary, for achieving objectives such as cost reduction, revenue maximization, and maximizing the use of renewable energy. Hence, the trading needs to be built on signal processing tools that can take such diverse set of constraints into consideration, and deliver an energy management solution, which ensures an efficient and robust operation of such heterogeneous and large-scale cyber-physical system. In this context,  considering the interactive and conflicting nature of energy trading, game theory is a very effective tool for modeling the decision-making process of the participants of such P2P networks. 

Essentially, game theory has been extensively used for design and analysis of the energy system, as we will see in the next section. However, due to the framework and aim of P2P energy trading, existing schemes might not be suitable to use in the same context. The is because: first, in P2P, the main objective would be to encourage the participants to trade energy with one another and thus comprise a community of energy without (or, very minimal) direct influence of the grid. Therefore, the price signal from the central power station may not affect the performance of the P2P trading as that influenced the scheduling and trading of energy in existing systems. Second, while the energy trading schemes in the smart grid have exploited various pricing schemes, including real-time and time-of-use pricing, P2P will necessitate the incorporation of more innovative pricing schemes. For example, being an independent decision maker, a prosumer may intend to sell its surplus energy at different rates to different buyers within the network. Hence, development of new pricing schemes would be necessary. Finally, relaxing the presence of centralized management from the trading scheme subsequently imposes a very high emphasis on the security of transactions of energy trading between the participants of the P2P network.

In this context, novel and innovative applications of game theoretic approaches will be necessary to design mechanisms for P2P energy trading. As such, this paper seeks to contribute towards achieving this goal by
\begin{itemize}
\item Presenting an overview of various games and auction theoretic approaches by following a systematic classification to provide information about the basic understanding and importance of game theory, and its extensive use in smart energy research. 
\item Focusing on the basic of P2P energy trading technique for integrating renewable energy sources into the grid through describing key features of such trading network, as well as well as by providing the explanation of an existing testbed that has deployed P2P trading for managing energy, and
\item Finally, zooming into the details of some specific game and auction theoretic models that have been used for P2P energy trading, and share some key results from those studies. This will provide the reader with an understanding of how to use game theory in P2P energy trading paradigm, and what is the potential benefit of it.
\end{itemize}

This paper can be used as a reference by both new and experienced researchers in an important emerging field in the smart grid research. Here it is important to note that signal processing is part of a much larger context within the smart grid context, and in this paper, we only overview the applications of various game and auction theoretic approaches for managing the trading between different entities within the energy network. Thus, this study seeks to complement existing game-theoretic literature in guiding engineers to effectively design and manage the substantial energy generated by DERs across the overall network without compromising the stability of the grid by enabling prosumers with conflicting interests to actively take part in energy trading with one another. For instance, in Table~\ref{table:TableArea}, we provide a summary of the pros and cons of game theory in addressing various challenges in energy system as well as its similar application in other sectors.

\begin{table}
\centering
\captionsetup{justification=centering}
\caption{A brief summary of advantages and limitations of using game theory for designing energy management schemes.}
\includegraphics[width=\columnwidth]{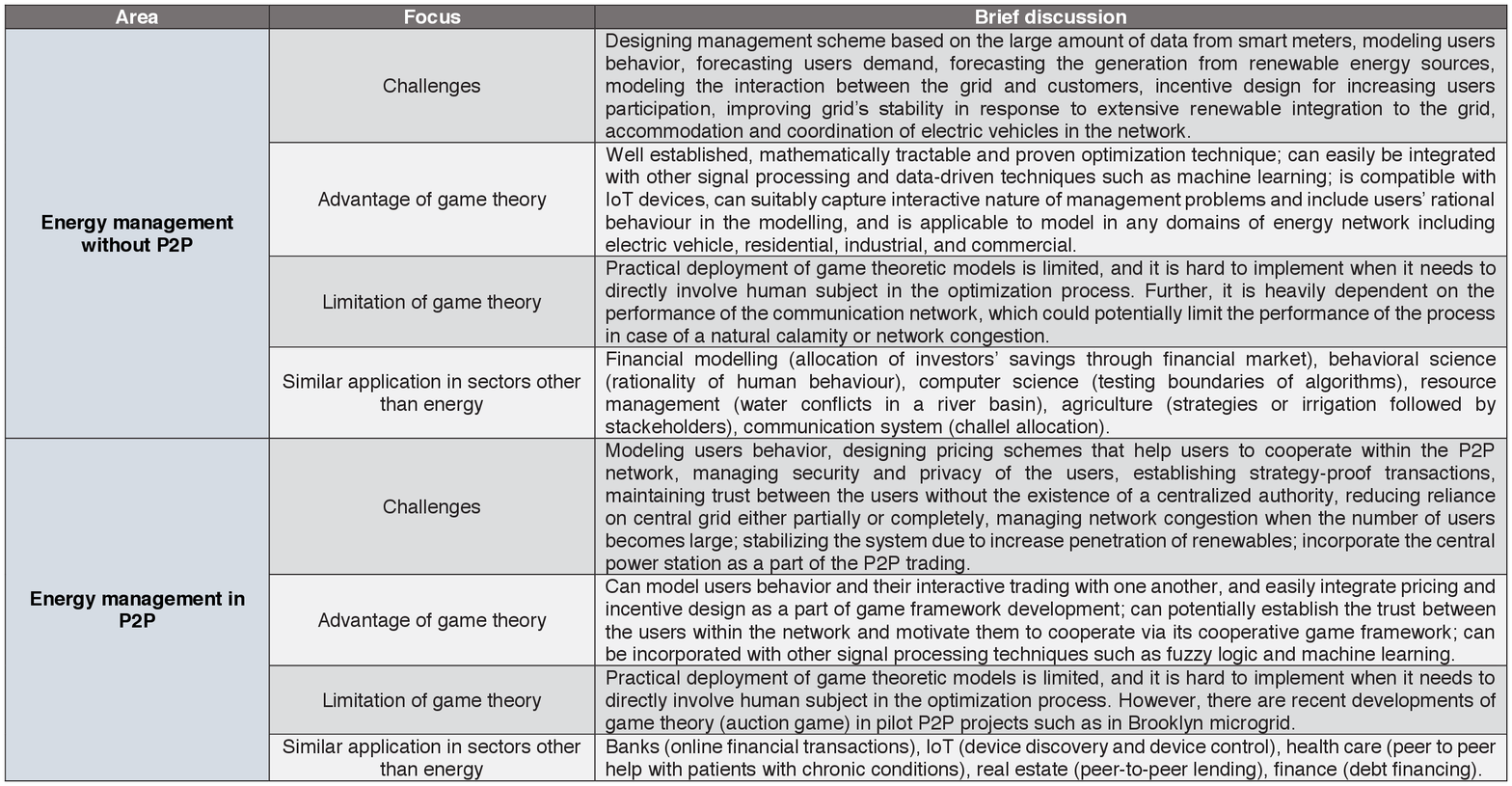}
\label{table:TableArea}
\end{table}

To this end, subsequent to a comprehensive literature review on game and auction theoretic approaches for smart grid energy management in Section~\ref{sec:state-of-the-art}, we explain the concept of P2P energy trading followed by a discussion on some key game and auction theoretic approaches that have been used for P2P energy trading in Section~\ref{sec:basic-concepts}. In Section~\ref{sec:outcomes}, we share some results from the game theoretic formulations and solution approaches explained in Section~\ref{sec:basic-concepts}, and compile the key insights from those results. Finally, Section~\ref{sec:conclusion} summarizes the entire discussion of the paper and makes some concluding remark on potential future research in P2P energy trading that can potentially be addressed with game and auction theoretic approaches.

\section{Game Theory for Smart Energy Management}\label{sec:state-of-the-art}
Within the context of energy management in smart grid, the applications of game and auction theoretic approaches are plentiful. On the one hand, non-cooperative games have been extensively used to schedule energy-related activities, and subsequently, trade the surplus energy with the buyers for making revenues. On the other hand, recently there has been large scale integration of alternative energy sources such as electric vehicles (EVs)~\cite{Wayes-J-TSG:2012}, solar photovoltaic (PV)~\cite{Tushar_TSG_2017}, and wind turbines~\cite{Marzband_IET_2016} into the grid that exploit game theory for efficiency energy trading to provide regulation services~\cite{Wu_TSG_2012} and efficient home energy management~\cite{Nguyen_JCN_2012}. In this section, we discuss the application of different game and auction theoretic approaches across EV domain, DER and storage domain, and service domain. To do so, we first provide a brief overview of the basic game theoretic concept. Then we provide a discussion of the different game and auction theoretic approaches that have been used to design various energy management schemes in these three domains for past few years. Here, it is important to note that the literature on game theory in energy management is extensive. Nonetheless, we keep our focus only on some of the key studies in each domain to comply with the journal's guideline on the total number of references.

\subsection{Basic Game Theoretic Concept}Game theory is a mathematical and signal processing tool~\cite{Bacci_SPM_2016} that analyzes strategies of competitive situations where the outcome of a participant's choice of action depends on the actions of other participants. It can be divided into two main branches including non-cooperative game theory and cooperative game theory. 

\subsubsection{Non-cooperative game} A non-cooperative game analyzes the strategic decision-making process of a number independent players that have partially or totally \emph{conflicting interest} over the outcome of a decision process, which is influenced by their actions. Such games allow players to take necessary action, e.g., optimal decision, without any coordination or communication. Here, it is important to note that the term \emph{non-cooperative} does not refer to the case that the players do not cooperate. Rather, it attributes to the fact that any cooperation that may arise in the non-cooperative game must be enforced with neither communication nor coordination of strategic choices among the players~\cite{Saad_GameSmartGrid_2012}. 

In general, the non-cooperative game can be divided into two categories: 1) static game, and 2) dynamic game.
\begin{itemize}
\item\emph{Static game:} In a static game, the players take their actions only once, either simultaneously or at different points in time. A static game can be defined in its \emph{strategic} form as $\{\mathcal{N},\left(\mathbf{S}_n\right)_{n\in\mathcal{N}},\left(U_n\right)_{n\in\mathcal{N}}\}$, where $\mathcal{N}$ is the set of all participating players in the game, and each player $n\in\mathcal{N}$ has a strategy set $\mathbf{S}_n$ from which it chooses an action $s_n\in\mathbf{S}_n$ to optimize its utility function $U_n$. The utility that a player $n$ attains is affected by choices of action $\mathbf{S}_{-n}$ of the players in set $\mathcal{N}\setminus\{n\}$.
\item\emph{Dynamic game:}  In contrast, players in a dynamic game act more than once and have some information regarding the choice of other players. In dynamic games, time plays a central role in the decision-making process of each player. Dynamic games can also be defined as that of a static game. Nonetheless, there is a need for some additional information including time and information set that are usually reflected in the utility functions.
\end{itemize}
For both static and dynamic non-cooperative games, the players may take their decisions either in a deterministic manner (pure strategies) or in a probabilistic manner (mixed strategies). 

The most popular solution concept of the non-cooperative game $\{\mathcal{N},\left(\mathbf{S}_n\right)_{n\in\mathcal{N}},\left(U_n\right)_{n\in\mathcal{N}}\}$ is the Nash equilibrium. A Nash equilibrium can be defined as a vector of actions $\mathbf{s}^*$ if and only if $U_n\left(\mathbf{s}^{*}\right)\geq U_n(s_n,\mathbf{s}_{-n}^{*})$, $\forall n\in\mathcal{N}$, where $\mathbf{s} = \left[s_n,\mathbf{s}_n\right]$. Thus, the Nash equilibrium refers to a stable state of a non-cooperative game, in which no player $n\in\mathcal{N}$ can improve its utility by unilaterally altering its action $s_n$ from $s_n^*$ when the actions of the other participating players $\mathcal{N}\setminus\{n\}$ are fixed at $\mathbf{s}_{-n}^{*}$. 
While a Nash equilibrium always exists in a non-cooperative game with mixed strategies, the existence is not guaranteed in a game with pure strategies. Further, a non-cooperative may also have multiple Nash equilibria, and in such cases, it is important to select an efficient and desirable Nash equilibrium as the solution of the game.   

\subsubsection{Cooperative game}\label{sec:BasicCooperativeGame} In cooperative games, on the other hand, the focus is on how one can provide incentives to independent decision makers to act together as one entity in order to improve their position in the game. In essence, both Nash bargaining and the coalitional game can be considered under the same umbrella of the cooperative game. Nash bargaining is the study of terms and conditions under which a number of players may agree to form a coalition. Meanwhile, coalitional games deal with the formation of coalitions~\cite{Saad_GameSmartGrid_2012}. In general, a coalition game can be expressed by the pair $\left(\mathcal{N}_c,\nu\right)$, which involves a set of players $\mathcal{N}_c$ who seek to form cooperative groups. $\nu$ is the value function associated with each coalition $\mathcal{S}\subseteq\mathcal{N}_c$ and is expressed by a real number to quantify the value of the respective coalition. The most common form of a coalitional game is the characteristic form~\cite{Saad_CoopGame_2009}, where the value of coalition is determined based on the members of that coalition, irrespective of how the players in the coalition are structured.  A coalitional game can be classified into three categories including 1) canonical coalitional game, 2) coalition formation game, and 3) coalitional graph games.
\begin{itemize}
\item\emph{Canonical coalitional game:} A canonical coalitional game can be expressed either as a game with transferable utility or as a game with non-transferrable utility. In this type of game, the formation of the grand coalition (the coalition of all players in the game) is never detrimental to the players, which pertains to the mathematical property of superadditivity. The main objectives of a canonical coalitional game are to study the properties and stability of the grand coalition, the gains resulting from the coalition, and the distribution of these gains in a fair manner to the players. The most renowned solution concept for the coalitional game is the core, which is directly related to the stability of grand coalition. Essentially the core is defined as the set of revenues $\mathbf{x}$ where no coalition $\mathcal{S}\subset\mathcal{N}_c$ has any incentive to reject the grand coalition for the proposed revenue allocation $\mathbf{x}$.
\item\emph{Coalition formation game:} In coalition formation game, network structure and cost for cooperation play a major role. In general, coalition formation game is not superadditive. Although forming coalition brings gains to its members, the gains are limited by a cost associated with coalition formation. As a consequence, the formation of a grand coalition is very rare in this type of game, and therefore the objective of static coalition formation game is to study the network coalitional structure. In dynamic coalition game, however, the coalitional game is subject to environmental changes including change in the number of players and variation in the network topology. Hence, the main objectives are to analyze the formation of a coalitional structure, through players' interaction, and then study the properties of the structure and its adaptability to environmental variations.
\item\emph{Coalition graph game:} Communication between players within a coalition plays a significant role in coalitional games. In fact, in some scenarios, the underlying communication structures between players can have a major impact on the utility and other characteristics of the coalitional game~\cite{Saad_CoopGame_2009}. The coalitional game that deals with such connectivity of communications between players is referred to as the coalitional graph game in \cite{Saad_CoopGame_2009}.  In this type of game, the main objectives are to derive low complexity distributed algorithms for players that wish to build a network graph (directed or undirected) and to study the properties (such as stability and efficiency) of the formed network graph.
\end{itemize}
\subsection{Energy Management in EV Domain}\label{sec:EVDomain} Recently, EVs have attracted much attention as a sustainable transport system not only for their environment-friendly features but also due to their capacities to assist the energy grid via vehicle-to-grid (V2G) and grid-to-vehicle (G2V) technologies. EV, as an independent entity, can participate in energy trading with other entities within the energy network either by coordinating its charging and discharging behavior in order to provide regulatory service to the grid~\cite{Wu_TSG_2012}, or by direct interaction with other traders within the network to decide on trading price and energy via negotiation~\cite{Wayes-J-TSG:2012}. In this context, now we provide a brief overview of some popular game theoretic approaches that have been studied in the literature to model energy trading by EVs.

As EV market is growing rapidly around the world, both the grid and EV owners will benefit if the flexible demand of EV charging can be properly managed~\cite{Liu_TSG_2017}. This has been done by devising new scheduling techniques for the charging and discharging of EV in \cite{Liu_TSG_2017,Nguyen_JCN_2012}, and \cite{Lee_TSG_2015} based on non-cooperative Nash game. For example, a day-ahead EV charging scheduling is proposed in \cite{Liu_TSG_2017} considering the impact of the electricity prices as well as the possible actions of other EVs. The unique Nash equilibrium is determined through quadratic programming, and the case studies are demonstrated using data from the Danish National Travel Surveys. Price competition between different EV charging stations with renewable power generators is studied in \cite{Lee_TSG_2015}, in which the authors show that the interaction between the EV charging stations can be captured via a supermodular game, which has a unique Nash equilibrium.  Finally, a smart charging and discharging process for multiple EVs is designed in \cite{Nguyen_JCN_2012} to optimize the energy consumption profile of a building.  In the non-cooperative energy charging and discharging scheduling game, the players are the EVs, and their strategies are the battery charging and discharging schedules, and the utility function of each EV is defined as the negative total energy payment to the building. Each EV independently selects its best strategy to maximize the utility function, and all EVs update the building planner with their energy charging and discharging schedules. It is shown that the EV owners will have incentives to participate in the proposed game. 

In the literature, auction-based game theoretic approaches, also known as auction games~\cite{Zou_TAC_2017}, have been used for studying the problem of coordination that arises from charging a population of EVs, bargaining of price at which energy is traded between an electricity market and different EVs~\cite{Wang_TSG_2014}, and for P2P electricity trading among EVs using newly introduced blockchain~\cite{Kang_TII_2017}. For instance, the authors in \cite{Zou_TAC_2017} use a progressive second price auction mechanism to ensure that incentive compatibility holds for the auction game, and the efficient bid profile of the auction game is achieved as the Nash equilibrium. In \cite{Wang_TSG_2014}, a non-cooperative game is formulated between storage units of EVs that are trading their stored energy. In the energy exchange market between the storage units and the smart grid elements, the price at which energy is traded is determined via an auction mechanism and is shown to admit at least one Nash equilibrium. An interesting blockchain based auction mechanism is developed in \cite{Kang_TII_2017}, in which the authors propose a consortium blockchain method to detail the operation of localized P2P energy trading. The electricity pricing and the amount of traded energy among the EVs are administered by an iterative double auction mechanism.

Another branch of game theory that has been exploited to design EV trading mechanism is the coalition game, which is essentially characterized by a set of players and a value function that quantifies the worth of a coalition. Examples of application of coalition game in EV energy trading can be found in \cite{Kumar_TDSC_2016} and \cite{Yu_ITJ_2014}. In \cite{Kumar_TDSC_2016}, the authors propose a Bayesian Coalition Negotiation Game as a means to perform energy management for EVs in the V2G environment. The game is used along with Learning Automata, wherein Learning Automata is stationed on EVs that are assumed as the players in the game. A Nash Equilibrium is shown to be achieved in the game using convergence theory. In \cite{Yu_ITJ_2014}, the authors argue that leveraging the cooperation among EVs can enable the grid to efficiently stimulate EV users to charge in load valley and discharge in load peak. As a consequence, the electricity load is well balanced, and the EV users also achieve a higher profit.  As such, the authors formulate the EV charging and discharging cooperation in the framework of a coalition game, and by doing so they show that the EV users have better satisfaction in the vehicle battery status and economic profit.

Hierarchical games, in particular, the Stackelberg game, is probably the most popular game that has been extensively used for designing energy trading mechanisms for EVs. For instance, in \cite{Wayes-J-TSG:2012}, the authors study a static non-cooperative Stackelberg game to facilitate energy trading between a smart grid and EV groups, which is then extended to a time-varying case that can incorporate and handle slowly varying environments. The energy trading between the aggregation of EVs and fast charging station is modeled as a Stackelberg game in \cite{Zhao_TSG_2017} to provide regulation reserves to the power system. In this study, EVs, as the followers of the game, can obtain a tradeoff between the benefits from energy consumption and reserves provision, by deciding their charging and reserve strategies. In \cite{Wang_TVT_2017}, a similar game is designed to capture the interaction between EVs and the charging system controller, and is shown that the game has a unique and optimal solution robust to poor communication channels. A two-stage Stackelberg game is studied in \cite{Yuan_TSG_2017} to address the problem of charging station pricing and EV charging station selection, in which the charging stations (leaders) announce their charging prices in stage I and the EVs (followers) make their selection of charging stations in stage II. It is shown that there always exists a unique charging station selection equilibrium in stage II, and such equilibrium depends on the charging stations' service capacities and the price difference between them. Similar examples of the hierarchical and other games in the EV domain can be found in \cite{Yang1_TVT_2016,Tan_TSG_2017,Mondol_IET_2015} and \cite{Zhu_Access_2016}. 
\subsection{Energy Management in DER and Storage Domain}\label{sec:DERDomain} The widespread adoption of DER in the power system can play a key role in creating a clean and reliable energy system with substantial environmental and other benefits. However, due to the fact that the energy production from these DERs is highly intermittent, their integration into the power system poses a significant challenge in maintaining the grid's stability. However, with suitable energy storage and energy management techniques, such intermittency can be addressed, and thus the benefit of using DERs can be increased significantly. As such, we discuss some of the game-theoretic techniques that have been used in the literature for effective energy trading in the DER and storage domain.

The development of two-way communication enables interaction between supply and demand side of the electricity network, and thus allows users to exploit Nash games to design energy management schemes for DERs. In \cite{Chen_TSG_2014}, for example, a game theoretic approach is analyzed to minimize the individual energy cost to consumers through scheduling their future energy consumption profiles. In particular, an instantaneous load billing scheme is designed to effectively convince the consumers to shift their peak-time consumption and to fairly charge the consumers for their energy purchase from the grid. With a view to reducing the cost of energy trading with the grid, a day-ahead optimization process regulated by an independent central unit is proposed in \cite{Atzeni_TSG_2013}. The existence of optimal strategies is proven, and further, the authors present a distributed algorithm to be run on the users' smart meters, which provides the optimal energy production and storage strategies, while preserving the privacy of the users and minimizing the required communication with the central unit. 

The auction game has been frequently used for trading both storage space and renewable energy from DERs. Example of such trading mechanism can be found in \cite{Cintuglu_TSG_2015} and \cite{Tushar_TSG_2016_sharing}. \cite{Cintuglu_TSG_2015} presents a real-time implementation of a multiagent-based game theory reverse auction model for microgrid market operations featuring conventional and renewable DERs. The proposed methodology was realistically implemented  in a smart grid system at the Florida International University. The investigation shows that the proposed algorithm and the industrial hardware-based infrastructure are suitable to implement in the existing electric utility grid. Meanwhile, the authors in \cite{Tushar_TSG_2016_sharing} utilize an auction game to study the solution of joint energy storage ownership sharing between multiple shared facility controllers and those dwelling in a residential community. It is shown that the auction process possesses both incentive compatibility and individual rationality properties, and is capable to enable the residential units to decide on the fraction of their energy storage capacity that they want to share with the shared facility controllers of the community to assist them in storing electricity. 

Recently, coalition games have also received attention for designing energy trading mechanism for users in residential areas that are equipped with DERs and storage devices. For example, in \cite{Lee_JSAC_2014}, the authors use a coalition game to study the cooperation between small-scale DERs and energy users to enable direct trading of energy without going through the retailers. The asymptotic Shapley value is shown to be in the core of the coalitional game such that \emph{no group} of small-scale DERs and energy users has an incentive to abandon the coalition, which implies the stable direct trading of energy for the proposed pricing scheme. Further, it is shown via numerical case studies that the scheme is suitable for practical implementation. The authors in \cite{Ni_IET_2016} focus on comprehensive economic power transaction of the multiple microgrids network with the multi-agent system and design a three-stage algorithm based on coalitional game strategy consisting of a request exchange stage, merge-and-split stage, and cooperative transaction stage. The developed algorithm enables microgrids to form coalitions, where each microgrid can exchange power directly by paying a transmission fee. 

Similar to EV domain, hierarchical games have also been extensively used for trading mechanisms in DER and storage domain\footnote{Nonetheless, due to the constraint on the total number references, we are unable to provide an overview of all of them.}. Two popular examples of such study include \cite{Maharjan_TSG_2013
} and \cite{Tushar-TIE:2014}. \cite{Maharjan_TSG_2013} proposes a distributed mechanism for energy trading among microgrids in a competitive market via a multileader-multi-follower Stackelberg game. The game is formulated between different utility companies and end-users to maximize the revenue of each utility company and the payoff of each user, where the existence of a unique Stackelberg equilibrium is proven. The paper also studies the impact of an attacker who can manipulate the price information from the utility companies and proposes a scheme based on the concept of shared reserve power to improve the grid reliability and ensure its dependability. A similar type of game is also designed in \cite{Tushar-TIE:2014}, in which the authors propose a three-party energy trading mechanism within a smart grid community. In particular, a non-cooperative Stackelberg game between the residential users and the shared facility controller is proposed to explore how both entities can benefit, in terms of achieved utility and reduction of total cost respectively, from their energy trading with each other and the grid. It is shown that the maximum benefit to the SFC, in terms of reduction in total cost, is obtained at the unique and strategy-proof Stackelberg equilibrium. Other games that have also been used in the DER and storage domain for energy management include dynamic games such as in \cite{Nava_TSMCS_2014}. 

\subsection{Energy Management in Service Domain}\label{sec:REGDomain} In this domain, we discuss how game theoretic approaches have been exploited to provide services to the grid and consumers via scheduling of energy-related activities by the users. Examples of such service include regulatory services such as voltage and frequency regulation, demand response regulation, services in terms of sharing of resources such as storage, and designing incentives for users. In this context, we can now explain how Nash game, auction game, coalition game and the hierarchical game has been used to provide these services to the energy users in EV, and DER and storage domain.

Nash game has been mostly used for providing demand response services to the grid. On the one hand, sometimes Nash game is exploited alone by the users to decide on the scheduling of their daily energy-related activities to participate in demand response, e.g., in \cite{Liu_TSG_2017} and \cite{Mohsenian-Rad_TSG_2010}. On the other hand, Nash game has also been played as a part of another game, such as hierarchical game, to reach the desired solution. For instance, in \cite{Wayes-J-TSG:2012} and \cite{Tushar-TIE:2014}, Nash game has been played by the followers, as part of the Stackelberg game, to reach an equilibrium solution. Nash game has also been exploited in \cite{Wu_TSG_2012}, in which the authors develop a smart pricing policy and design a mechanism to achieve optimal frequency regulation performance in a distributed fashion.

The application of auction games can be found in designing services like storage sharing~\cite{Tushar_TSG_2016_sharing}, demand response~\cite{Wang_TSG_2014}, and frequency regulation \cite{Cintuglu_TII_2017}. For example, in \cite{Cintuglu_TII_2017}, the authors present a bidding behavior modeling and an auction architecture consisting of a central aggregator and networked microgrid agents. The bidding behavioral states of the microgrid agents are formalized for the belief updates and short-term policy determination to maximize the individual profit. Then, a reverse auction model is adapted to enable competitive negotiations between the central aggregator and networked microgrid agents. The auction and aggregation processes were implemented in a power system control area to contribute to frequency control. Further, auction games such as in \cite{Tushar_TSG_2016_sharing} have also been exploited by incentivizing users to participate in energy management.

Coalition games, including both coalition formation and a canonical coalition, has applied for designing services in the energy sector. For instance, demand response regulation in the EV domain has been implemented by using a coalition formation game in \cite{Kumar_TDSC_2016}. In \cite{Lee_JSAC_2014}, the authors demonstrate how to incentivize energy users with small-scale energy power production unit such as rooftop solar panels to directly energy trade with other users within a community instead of trading with the retailer.  Further, the exploration of coalition game for regulation service can be found in \cite{Yu_ITJ_2014}, in which the authors design a coalition formation game to schedule the charging and discharging of EVs within a smart grid network such that the grid's stability is not compromised.

Hierarchical games have covered almost all aspects of service domain. For example, demand response regulation has been covered in \cite{Maharjan_TSG_2013}, whereas the authors in \cite{Tushar-TIE:2014} show how hierarchical game can influence users with suitable incentives to participate in energy trading. In \cite{Tan_TSG_2017}, a hierarchical game is proposed to provide frequency regulation under a V2G scenario. In particular, a hierarchical Markov game is designed to coordinate the charging process of EVs. It is shown that the Markov game optimizes the regulation capacity of the aggregator, and thus strengthen its ability in bidding a more favorable frequency regulation price. Regulation service with a hierarchical game is also implemented in \cite{Zhao_TSG_2017}. Further, application of Stackelberg game to decide on a price for sharing energy storage device can be found in \cite{Tushar_TSG_2016_sharing}. Finally, another example of the game-theoretic approach in this domain can be found in \cite{Zhu_Conf_2011}.

\begin{table}
\centering
\captionsetup{justification=centering}
\caption{This table demonstrates the importance and extensive use of game theory in smart energy domain.}
\includegraphics[width=\textwidth]{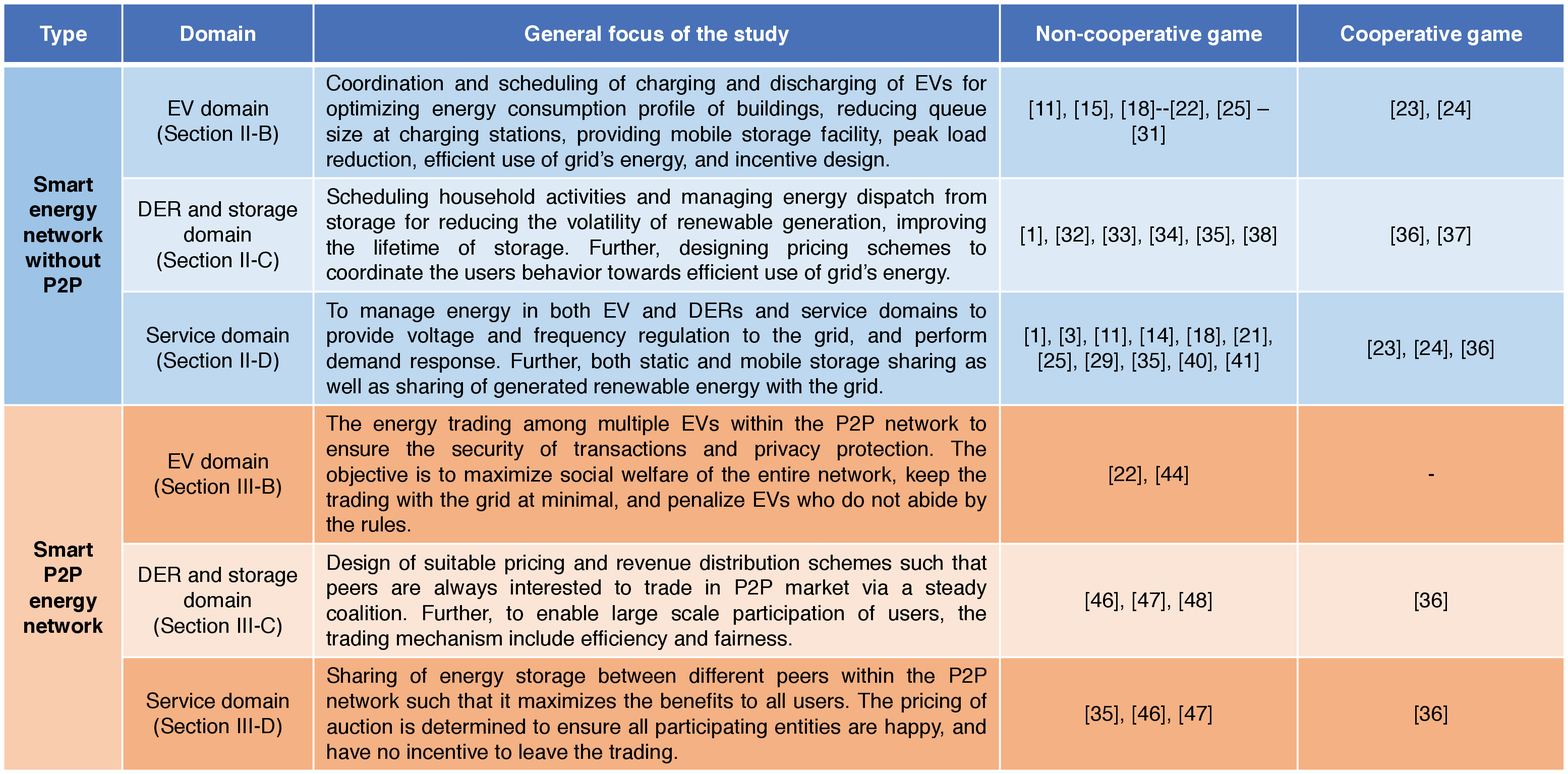}
\label{table:TableSummery}
\end{table}

As can be seen from the above discussion, which is also summarized in Table~\ref{table:TableSummery}, the application of game theory in the paradigm of energy trading and management is extensive. However, the discussion on its application in the field of P2P energy trading is limited, which could be due to the recent emergence and exploration of P2P trading framework in the energy domain. In this context, what follows is an introduction to the P2P energy network followed by a discussion on some specific game theoretic techniques that have been used for designing P2P energy trading.

\section{Game Theory for Energy Management in P2P Network}\label{sec:basic-concepts}
\subsection{P2P Energy Network}P2P is a network, in which the members or peers of the network share a part of their own resources and information to facilitate certain applications. Each peer is both a provider and receiver of the resource and can directly communicate with rest of the peers of the network without the intervention of any intermediate node~\cite{Giotitsas_2015}. This enables the network to be resilient against the failure of one or more peers within the network, and continue to operate normally. Thus, new peers can be added or old one can be replaced without altering the operational structure of the system. As such, P2P energy network, as shown in Fig.~\ref{fig:figure_P2P}, consists of a number of energy users including both consumers and prosumers. Prosumers are equipped with small-scale DER units such as rooftop solar panels and small wind turbines. The production of energy takes place within each house or nearby to reduce transmission losses and utilize cogeneration, if possible. When a prosumer has surplus energy, it can either store this energy within its storage device, if there is any, or distribute the energy among other peers within the network to avoid having wasted energy~\cite{Giotitsas_2015}. As such, empowers the users of the energy network to take control of the production and consumption of energy within the community without any central control authority such as the grid~\cite{Tushar_TSG_2016_sharing} by potentially avoid employing complex algorithms and technological equipment to negotiate pricing for buying and selling of energy and storage~\cite{Giotitsas_2015}.

A P2P energy network consists of two main components including the virtual energy trading platform, and the physical energy network~\cite{Mengelkamp_AE_2017}. \emph{Virtual energy trading platform} provides the technical infrastructure for the local electricity market. It has to be based on a secured information system, e.g., blockchain based architecture in Brooklyn microgrid, in which the transfer of all kinds of information takes place. It needs to be implemented in a way such that each peer has equal access to avoid discrimination. For example, the generation, demand, and consumption data of a peer are transferred from its smart meter to the virtual layer over a secured communication network. Then, buy and sell orders are created in the virtual layer based on this information from the smart meter, which is then sent to the appropriate market mechanism to facilitate energy trading. Once the matching of buy and sell orders are completed between different peers, the payment is carried out, and subsequently, the exchange of energy takes place over the physical layer. 

On the other hand, \emph{physical energy network} is the distribution grid, which is used for the physical transfer of energy among the peers. This physical network could be the traditional distributed grid network provided and maintained by the independent system operator (ISO). Alternatively, it can be an additional separate physical microgrid distribution grid in conjunction with the traditional grid, which provides the peers of the network with the flexibility to be physically disconnected from the main grid in case of an emergency event~\cite{Mengelkamp_AE_2017}. Here it is important to note that the financial transactions that are carried out between different peers in the virtual platform have no influence on the physical delivery of electricity. Rather, the payment can be thought of as the payment from the consumers to their producing prosumers within the P2P network for feeding the renewable generation into the distribution grid~\cite{Mengelkamp_AE_2017}.

\subsubsection{Key features}According to \cite{Mengelkamp_AE_2017} and \cite{Llic_Conf_2012}, an energy network should have seven key features, as explained in the following sections, for successful P2P energy trading.

\emph{Market participants:} A clear definition of market participants, as well as the purpose of the P2P energy trading, must be established, and the form of energy that is traded in the market should be clarified. P2P energy trading necessitates the existence of a sufficient number of market participants within the network, and a subgroup of the participants need to have the capacity to produce energy. The purpose of P2P energy trading, e.g., increasing the use of renewable energy or reducing dependency on the main grid, affects the design of pricing scheme and market mechanism of the trading market. Further, the form of energy traded in the market should be defined, i.e., whether the energy is traded in form of electricity, heat or combination of both.

\emph{Grid connection:} For balancing the energy generation and consumption within the P2P energy trading network, it is imperative that the connection points towards the main grid are well defined. At these connection points, it is possible to connect a smart meter to evaluate the performance of the P2P energy network, e.g., how much energy cost the participants can save by not buying from the grid. If a physical microgrid distribution network exists between the participants, it swiftly decouple itself from the main grid in case of an emergency. However, for such island-mode operation, participants should have enough generation capacity to ensure the appropriate level of supply security and resiliency.  Nonetheless, if the P2P energy trading is only  conducted over the existing traditional distribution network, such island-mode operation is not possible.

\emph{Information system:} A high performing information system is the heart of any P2P energy trading network. Such an information system is necessary for: 1) connecting all market participants for energy trading, 2) providing the participants with a suitable market platform, 3) to render the participants with access to the market, and 4) to monitor the market operation. It is important that every market participant has equal access to the market information without any discrepancy. An example of such an information system is the blockchain based smart contracts~\cite{Kang_TII_2017}.

\emph{Market operation:} Market operation of P2P energy trading is facilitated by the information system. It consists of the market's allocation, payment rules, and a clearly defined bidding format. The main purpose is to provide an efficient energy trading experience by matching the market participants' sell and buy orders in near real-time granularity. In market operation, the constraint of energy generation influences the thresholds of maximum and a minimum allocation of energy. Different market time horizon can exist in the market operation, such as day ahead and intraday, to cover various stages of the electricity market, and the market operation should be able to produce efficient allocation in every stage.

\emph{Pricing mechanism:} The objective of pricing mechanism is to efficiently balance energy supply and demand, and is implemented as a part of the market operation. Examples of pricing mechanisms include auctions with individual or uniform clearing price. Pricing mechanism for P2P energy trading has a big difference with that of the traditional energy market. With traditional energy, a large part of energy price consists of taxes and surcharges, whereas in a P2P trading market such tax and surcharges are absent due to zero marginal cost of renewable energy. Nevertheless, pricing needs to reflect the state of energy within the P2P energy network. For example, a higher surplus should lower the price of P2P energy trading and vice versa.

\emph{Automatic energy management system:} The purpose of automatic energy management system (AEMS) is to secure the supply of energy for a market participant while implementing a specific bidding strategy. To do so, AEMS has access to the real-time demand and supply information of its market participant, and based on these data, an AEMS forecasts the generation and consumption profile as well as  develops the bidding strategy. The AEMS of a rational user would always buy energy at the microgrid market when the price falls below its maximum price limit. Nonetheless, individual agents' intelligent bidding strategies shall employ varying prices at different times and are expected to be one of the core components of active P2P energy markets in the future.

\emph{Regulation:} Finally, the regulation is the feature of a P2P energy trading that determines how such markets fit into the current energy policy. That is, government rules decide which market design is allowed, how taxes and fees are distributed, and in which way the market is integrated into the traditional energy market and energy supply system. Hence, governments either can support P2P energy markets to accelerate the efficient utilization of renewable energy resources and to decrease environmental degeneration by regulatory changes, or discourage the implementation of such markets if these result in negative impacts on the current traditional energy system.

\subsubsection{Brooklyn TransActive P2P project}Now we focus on an existing pilot project on P2P energy trading, which is built in Brooklyn, New York. This discussion on a real P2P energy network will provide the reader with a brief idea of how the P2P energy trading is being envisioned to be conducted in the future energy market\footnote{As of today, local P2P energy trading without any utility involvement is yet to be covered by the regulation, which decides how such market fits into the current energy policy~\cite{Mengelkamp_AE_2017}.}. The choice of Brooklyn microgrid for this discussion is motivated by the extensivity of the project as well as the successful implementation of trading techniques as portrayed by their recent pilot demonstration.
\begin{figure}
\centering
\captionsetup{justification=centering}
\includegraphics[width=\textwidth]{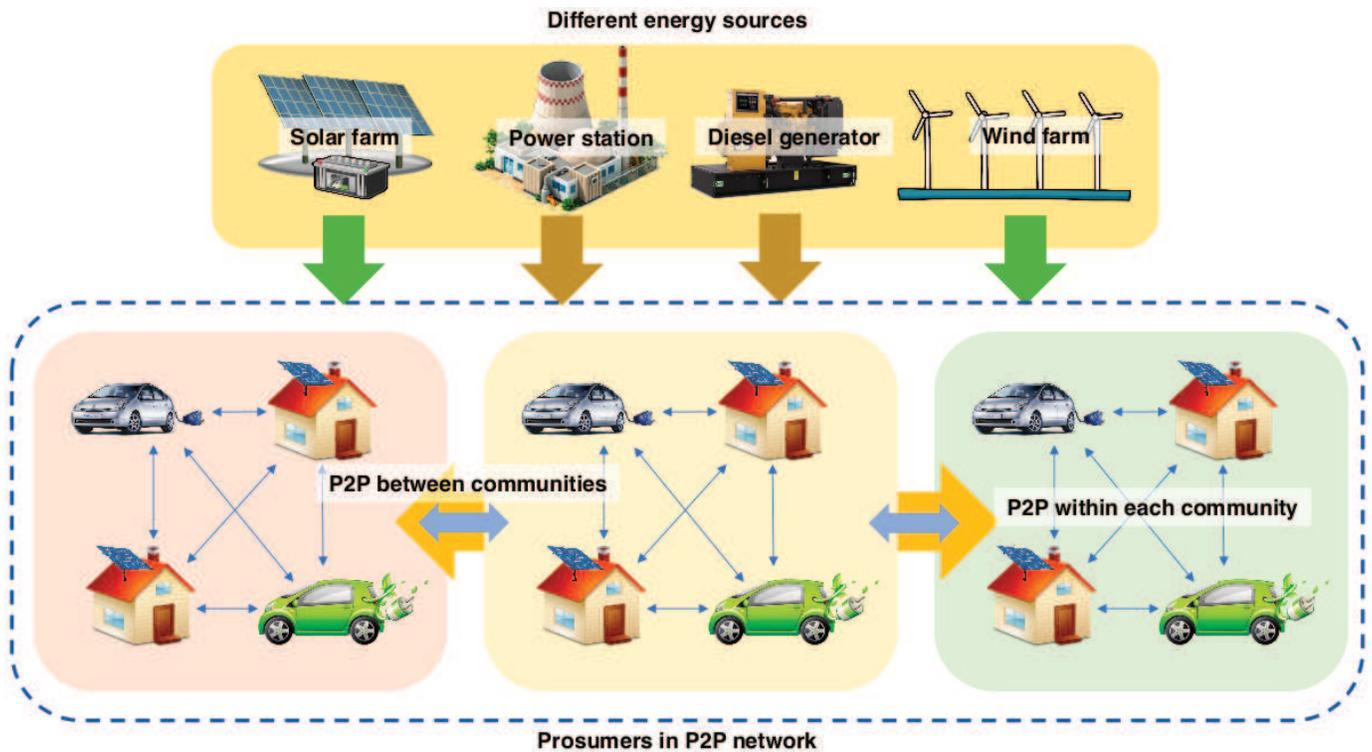}
\caption{This figure demonstrates the topology of Brooklyn microgrid, and is inspired from \cite{Mengelkamp_AE_2017}.}
\label{fig:BrooklynMicrogrid}
\end{figure}
Brooklyn microgrid project, which we will refer to as BMP for the rest of the paper, consists of a microgrid market in Brooklyn, New York. The project is run by LO3 Energy, and the participants of the BMP are located across three distribution grid including Borough Hall, the Park Slope, and the Bay Ridge. As shown in Fig.~\ref{fig:BrooklynMicrogrid}, the BMP trading network consists of a physical layer and a virtual layer. In the physical layer, the BMP uses the traditional grid to supply physical energy flow. However, it also has a physical microgrid network among a limited number of housing blocks\footnote{Which in particular comprises 10-by-10 housing blocks at present.} that can be decoupled from the main grid in case of emergency. The virtual layer is completely separated from the physical layer. The virtual layer is implemented on top of the existing physical grid infrastructure, and provides the technical infrastructure for the local electricity market, and is based on Tendermint protocol based private blockchain called TransActive blockchain architecture~\cite{Mengelkamp_AE_2017}. Each peer must have a blockchain account to participate in the P2P energy trading. A TransActive meter is installed within the house of each peer that communicate with his blockchain account and transfers energy generation and demand data from the TransActive meter.
\begin{figure}
\centering
\captionsetup{justification=centering}
\includegraphics[width=0.8\textwidth]{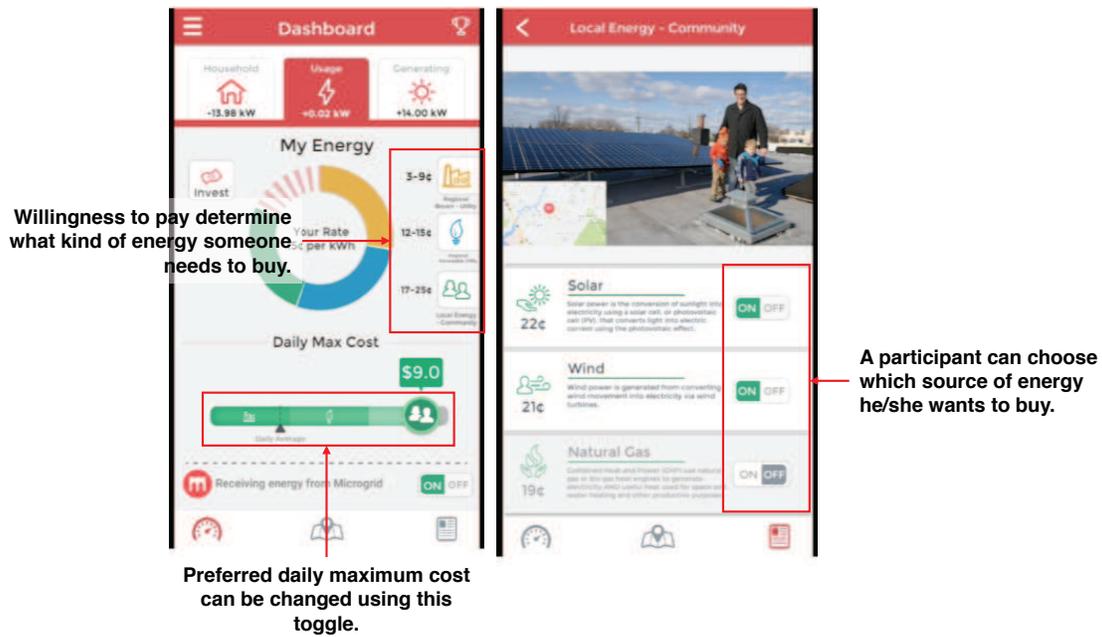}
\caption{This figure demonstrates the screen shot of the mobile application used in the BMP. The screen shot of the application is taken from the Brooklyn microgrid website: https://www.brooklyn.energy/video-gallery.}
\label{fig:MobileApp}
\end{figure}

The energy trading at BMP is mostly done automatically by an AEMS, and only requires several preferences from its market participants. In particular, the participants use a mobile app (name BMG App), through which they can choose their preferences on the source of energy and price limits for the AEMS to conduct the energy trading. An example diagram of the mobile application\footnote{LO3 energy has given permission to use this screen shot of the mobile application in this paper.} is shown in Fig.~\ref{fig:MobileApp}. Although the participants can change their preferences at any time, it is also possible for the participants to choose one preference, and set that for all the time without any further interaction with the mobile application. Now, once the preferences are submitted through the mobile application by the participants, the energy trading between two participants (one consumer and one prosumer) takes place following a step-by-step process explained below~\cite{Mengelkamp_AE_2017}:
\begin{itemize}
\item[]\emph{\textbf{Step 1:}} The buy and sell orders of a consumer and a prosumer are submitted to the market by their AEMSs respectively. Any buy or sell order consists of a quantity and a price.
\item[]\emph{\textbf{Step 2:}} The market mechanism is a closed order book with a time discrete double auction in $15$ minutes time slot. In the double auction: 1) Consumers constantly bid their maximum price limit for their preferred energy sources, 2) Prosumers bid the minimum price limit that they request for selling their generation on the market, 3) The highest bidder is allocated first, and then lower bidders are allocated following a merit-order dispatch, and finally 4) The last allocated bid price represents the market clearing price for that particular time slot. 
\item[]\emph{\textbf{Step 3:}} Consumers that cannot undercut the clearing price are supplied by additional sources.
\item[]\emph{\textbf{Step 4:}} Financial transactions are carried out between the allocated market participants of that particular time slot according to predefined payment rules.
\item[]\emph{\textbf{Step 5:}} Local trading is then realized in the virtual layer, and the transfer of funds is completed.
\item[]\emph{\textbf{Step 6:}} In the physical, upon completion of the transaction of payment, the prosumers feed their renewable generation into the distribution grid\footnote{Prosumers may need to pay a subscription fee to the utility grid to use its network for P2P trading of energy.} for the consumers to consume.
\end{itemize}
A more comprehensive discussion on the Brooklyn microgrid can be found in \cite{Mengelkamp_AE_2017}.

As we now have some idea of how a P2P energy network works in real life, we detail some specific game theoretic approaches that have been used for P2P energy trading in EV, DER and storage, and service domains. For each domain, we take a specific study as an example and then explain in detail how the relevant game of that particular study is used to design the P2P trading scheme. Note that such explanation would help the reader to visualize how they may use such approaches to design games between different nodes in a P2P networks to attain various energy management objectives. In particular, we detail an auction game for EV domain, a coalition game for DER and storage domain, and a hybrid game (auction and Stackelberg game together) for service domain in the next section. Some interesting results from these study will be shared in Section~\ref{sec:outcomes}.

\subsection{P2P energy trading in EV domain} Recently, P2P energy trading in EV domain is gaining much attention, and the studies are being conducted based on both different game theoretic and optimization approaches, such as in \cite{Kang_TII_2017,Saad_SmartGridComm_2011} and \cite{Alvaro-Hermana_TSM_2016} respectively. In this section, however, we keep our focus on the study in \cite{Kang_TII_2017}, where an interesting exploration of \emph{auction game} for P2P energy trading can be found in the EV domain. In particular, this study designs a P2P energy trading technique among EVs in the smart grid via auction game, and ensure the security and privacy of the transactions by incorporating a consortium blockchain within the trading mechanism. Note that to explore how auction game is used in \cite{Kang_TII_2017} for P2P energy trading in the EV domain, we will only focus on the use of an auction game for P2P trading and ignore the design of consortium blockchain in the discussion. Interested reader can find the detail of the consortium blockchain in \cite{Kang_TII_2017}. As such, first we note that the main objective of all EVs within the P2P energy network is to maximize the social welfare, and the model for this localized P2P energy trading consists of three main components:
\begin{itemize}
\item EVs: The EVs play different roles in the proposed P2P electricity trading at charging stations: charging EVs, discharging EVs, and idle EVs. Each EV can choose its own role according to its current energy state and driving plan.
\item Local aggregator: Local aggregators are the energy brokers that provide access points for electricity and wireless communication services for EVs. Each charging EV sends a request for electricity demand to the nearest local aggregator. Then, the energy broker does a statistics of local electricity demand and announces this demand to local EVs. In response, EVs with surplus electricity submit selling prices to the broker. The energy broker acts as an auctioneer to carry out an iterative double auction among EVs and matches electricity trading pairs of EVs.
\item Smart meter: Each charging pole is equipped with the smart meter that calculates and records the amount of traded electricity in real time. The records in the smart meters are used by the charging EVs to pay the discharging EVs.
\end{itemize}

In each charging station, a local aggregator can communicate with any local EV to establish a real-time electricity trading market and facilitate electricity trading between any charging EV and any discharging EV in the network. Each charging EV $CV_i^n$, which is connected to a local aggregator $n$ has a particular energy demand $c_{ij}^n$ from the discharging EV $DV_j^n$ connected to the same local aggregator. Meanwhile, $d_{ji}^n$ is the amount of energy that a discharging EV $DV_j^n$ supplies to $CV_i^n$ in local aggregator $n$. Now, due to the charging and discharging of $c_{ij}^n$ and $d_{ji}^n$, the satisfaction and cost function of charging and discharging EVs are respectively given by~\cite{Kang_TII_2017}:
\begin{equation}
U_i(\mathbf{C}_i^n) = w_i\ln(\eta\sum_{j=1}^J c_{ij}^n-c_i^{n,\text{min}}+1),\label{eqn:equation-1}
\end{equation}
and
\begin{equation}
L_j(\mathbf{D}_j^n)=l_1\sum_{i=1}^I(d_{ji}^n)^2 + l_2\sum_{i=1}^I d_{ji}^n.\label{eqn:equation-2}
\end{equation}
In \eqref{eqn:equation-1}, $\eta$ is average charging efficiency from discharging EVs to $CV_i^n$, and $w_i$ is the charging willingness of $CV_i^n$. In \eqref{eqn:equation-2}, $l_1$ and $l_2$ are cost factors. Please note that these satisfaction and cost function refer to the benefit and cost, in terms of real numbers, that each charging and discharging EV can obtain by participating in P2P energy trading with one another and may vary with the changes in the parameters such as energy price, charging willingness, and cost factors according to \eqref{eqn:equation-1} and \eqref{eqn:equation-2}. 

Nonetheless, as mentioned earlier, the purpose of this P2P trading is to maximize the social welfare. This is done by the local aggregator $n$ by interacting with both charging and discharging EVs to decide on the suitable charging and discharging energy vector $\mathbf{C}^n$ and $\mathbf{D}^n$ for trading. In doing so, as explained in \cite{Kang_TII_2017}, the local aggregator $n$ that is working as an energy broker not only to meet the demand of charging EVs but also maximize electricity allocation efficiency. As such, the overall objective function of the social welfare problem becomes the difference of \eqref{eqn:equation-1} and \eqref{eqn:equation-2} for the participating EV.  Note that for the social welfare maximization problem, it is necessary that the energy broker obtains true and complete information of all EVs' utility and cost functions. The complete information of EVs includes current energy state, battery capacity and so on. However, this is private information for EVs that EVs may not be willing to share with the energy broker. To address the issue, the designed mechanism needs to extract hidden information from the EVs. 

\emph{Auction mechanism:} In this context, an auction game, which is a part of the non-cooperative game, is efficient to elicit the hidden information in a real and competitive energy market, and therefore used in \cite{Kang_TII_2017} to facilitate P2P energy trading among the EVs. A double auction technique possesses the individually rational and weakly budget balanced properties, which confirm that the participating EVs bid truthfully according to privacy information, and at the same time the energy broker would not lose money to conduct the auction, respectively. In this context, the auction game in  \cite{Kang_TII_2017} is adopted by following an iterative step-by-step manner as mentioned below. In each iteration,
\begin{itemize}
\item[]\emph{\textbf{Step 1:}} Each participating charging and discharging EV submit its bid price to the auctioneer.
\item[]\emph{\textbf{Step 2:}} Based on the received bid price vector of buying energy (vector of bid prices from all charging EVs) and bid price vector of selling energy (vector of bid prices from all discharging EVs), the auctioneer produces optimal allocation of demand and supply energy vectors by following a pre-defined allocation policy, and broadcasts them to the participating EVs. 
\item[]\emph{\textbf{Step 3:}} According to the received allocated energy vectors from the auctioneer, each EV determines its optimal bid price, i.e., the optimal bid price for selling by discharging EV and optimal bid price for buying by charging EV.
\item[]\emph{\textbf{Step 4:}} Each EV submits its optimal bid price to the auctioneer.
\item[]\emph{\textbf{Step 5:}} The auctioneer receives the vectors of optimal bid price from both types of EVs and benchmark against pre-defined criteria to understand whether the optimal solution is obtained.
\item[]\emph{\textbf{Step 6:}} If optimal solution is obtained, the auction game is completed, and no further iteration is needed. Otherwise, the process re-iterate from Step 2 again.
\end{itemize}

In the proposed auction process, the auctioneer monitors localized P2P energy trading in real time. When some unexpected incidents happen, for example, few EVs may leave suddenly from the scheduled trades, the auctioneer may restart the auction process again, and a new energy trading process is executed. However, in such cases, the abruptly disconnected EVs are held accountable and are made to pay a penalty of disconnection. As shown in \cite{Kang_TII_2017}, the considered auction-based approach can obtain efficient energy allocation solution with optimal social welfare in the energy market, without requiring the participants to share complete private information about their satisfaction and cost functions.

\subsection{P2P Energy Trading in DER and Storage Domain} To study the application of game theoretic approach for P2P energy trading in the DER and storage domain, we will focus on the study in \cite{Lee_JSAC_2014}. In this study, the authors design a coalition game to enable direct energy trading from one peer to another peer within the energy network. To do so, the customers within the network are divided into two kinds. The first type of customers are small-scale electricity suppliers, who have renewable energy facilities (such as houses with rooftop solar panels), and can sell their excess energy to the market for monetary profit. Another type of customers is end users who need to buy energy to conduct energy-related activities. The amount of electricity supply and energy demand vary across time and may differ for each entity. While the customers can trade their respective energy amount in the traditional market with retailers, in \cite{Lee_JSAC_2014}, the authors show that energy trading in the P2P market could be more beneficial for both the end users and small-scale electricity suppliers. This is mainly due to the fact that there is a significant difference between the wholesale price $p_\text{wp}$ (selling price per unit of energy) and retail price $p_\text{rp}$ (purchase price per unit of energy) in the traditional electricity market, and $p_\text{rp}>p_\text{wp}$. Hence, the monetary benefit that a customer may gain in terms of either obtaining revenue or to reduce cost is very low. In P2P energy trading, on the other hand, the trading price $p_\text{p2p}$ is set between the wholesale price and the retail price, i.e., $p_\text{wp}\leq p_\text{p2p}\leq p_\text{rp}$. In \cite{Lee_JSAC_2014}, it is shown that such a choice of price is beneficial to both the small-scale sellers and end-users.

The coalition game formed between the small-scale electricity suppliers and end-users is a canonical coalition game with transferable utility, and the authors determine the price $p_\text{p2p}$ for P2P energy trading based on asymptotic shapely value~\cite{Saad_CoopGame_2009}. The canonical coalition game is formally defined by the pair $(\mathcal{N}_c,\nu)$, where $\mathcal{N}_c$ is the union of the set $\mathcal{N}_s$ of small-scale electricity suppliers and the set $\mathcal{N}_u$ of end-users, and as described in Section~\ref{sec:BasicCooperativeGame}, $\nu$ is a real number that refers to the total benefit that all participants of the game attain for forming the coalition. It is considered that the value function $\nu$ depends on the net surplus and deficient energy of the coalition. That is all participants of the coalition primarily trade their energy among themselves with a price $p_\text{p2p}$. Then, if there is any net surplus from the coalition, it is sold in the retail market at a rate of $p_\text{wp}$ per unit of energy, and buy energy at a price $p_\text{rp}$ per unit if there is any net deficiency. Thus, $\nu$ = ($p_\text{wp}\times$ net surplus) - ($p_\text{wp}\times$ net deficiency).

Now, to effectively perform P2P energy trading, a coalition needs to satisfy three properties as explained below.
\begin{itemize}
\item\emph{Superadditivity:} Formation of grant coalition needs to be beneficial for all participating customers of the coalition. In other words, it is always beneficial for the small-scale electricity suppliers and the end-users to trade in P2P energy trading, rather than trade in the traditional market. Thus, both parties are interested to maximize the total revenue of the coalition. To meet this property, however, the value function $\nu$ needs to be superadditive, which is the case in \cite{Lee_JSAC_2014}. Superadditivity refers to the condition that the total benefit that set of small-scale electricity suppliers and end-users obtained by forming the grand coalition is at least equal to the total benefit that they achieve by trading separately. 
\item\emph{Core:} There should be a fair distribution of total revenue among each customers forming the coalition. In P2P energy trading, this allocation of revenue can be done by suitably adjusting the trading price $p_\text{p2p}$ such that no subgroup of customers can obtain more revenue by deviating from the P2P trading. The feasible allocation of such revenue among participants is known as the core of a coalition, and if the core of a coalition is non-empty, no group of users has any incentive to leave that coalition. It is shown in \cite{Lee_JSAC_2014}, for the considered study, there is a non-empty core for the coalition when $p_\text{rp}>p_\text{wp}$.
\item\emph{Stability:} When all customers receive their respective revenues, which is at the core, no one wants to leave the coalition, which makes the coalition stable. In other words, all customers within the network continue to participate in P2P energy trading among themselves.
\end{itemize}
Nevertheless, derivation of fair distribution of revenue is complex and could be computationally expensive. There are a number of mechanisms that can be used in literature to determine fair revenue distribution such as Shapley value, nucleolus, and proportional fairness. In this work, $p_\text{p2p}$ is derived according to the Shapley value. Essentially, the concept of Shapley value is based on three axioms: efficiency, symmetry, and balanced contribution, and is a measure of the contribution made by each customer participating in P2P energy trading. By allocating revenue to each customer according to its Shapley value, the revenue of the P2P energy trading is fairly divided. This is due to the fact that what each customer obtains corresponds to its contribution to P2P energy trading. 

However, as the number of customers within coalition becomes very large, the number of computations increases prohibitively to determine the Shapley value of each customer. As such, in \cite{Lee_JSAC_2014}, the revenue distribution is conducted using an asymptotic Shapley value. For detail on the derivation of asymptotic Shapley value, please see \cite{Lee_JSAC_2014}. It is because the derived asymptotic Shapley value lies within the core of the coalition game. As a consequence, according to the third property mentioned above, the coalition is stable even for a very large number of customers. In other words, the proposed P2P energy trading scheme is suitable to adopt in an energy network consisting of a very large number of customers.

Based on the above discussion, the adapted canonical coalition game to design P2P energy trading in DER and storage domain can be summarized in following steps:
\begin{itemize}
\item[]\emph{\textbf{Step 1:}} Choose or design a system model suitable to incorporate P2P energy trading.
\item[]\emph{\textbf{Step 2:}} Design a value function that captures the benefit of the coalition. Determine whether the value function possesses the property of superadditivity.
\item[]\emph{\textbf{Step 3:}} Investigate the existence of the core in the coalition game.
\item[]\emph{\textbf{Step 4:}} If the core is non-empty, design a suitable revenue distribution mechanism that lies at the core. Thus, the coalition is stable.
\item[]\emph{\textbf{Step 5:}} Ensure that the design of revenue distribution technique can accommodate a large number of customers. This will confirm the practicality of actual implementation of the model.
\end{itemize} 
Other game theoretic application for P2P trading in the DER and storage domain can be found in \cite{Samadi_TSG_2016,Kalathil_TSG_2017} and \cite{Zhang_EP_2016}.
\begin{figure}
\centering
\captionsetup{justification=centering}
\includegraphics[width=0.8\textwidth]{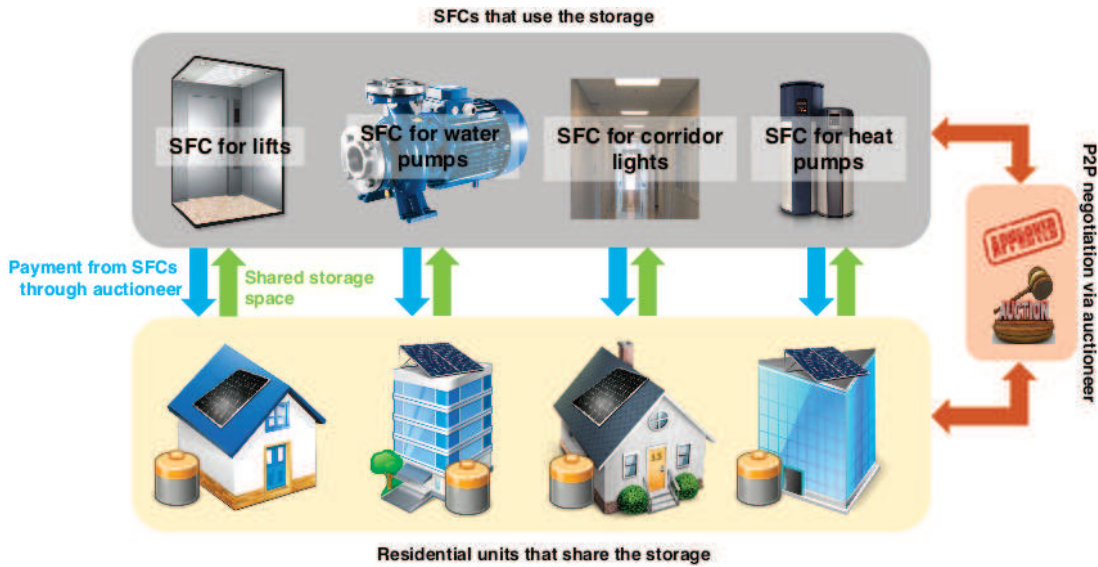}
\caption{Demonstration of the system model of application of P2P energy trading in service domain.}
\label{fig:ServiceDomainAuctionGame}
\end{figure}
\subsection{P2P Energy Trading in Service Domain}\label{sec:P2PServiceDomain} Finally, we discuss the application of game theory in P2P energy trading for service domain by describing the game proposed in \cite{Tushar_TSG_2016_sharing}. In \cite{Tushar_TSG_2016_sharing}, the authors propose an interesting integration of auction game with Stackelberg game to provide demand response service to the users of the network by sharing of energy storage device. In particular, the paper studies the solution of a joint energy storage sharing between multiple residential units and shared facility controllers (SFCs) within a community by enabling the residential units to decide on the fraction of their ES capacity that they may share with the SFCs of the community in order to assist them in storing electricity, e.g., for fulfilling the demand of various shared facilities. To do so, a modified auction game is designed that captures the interaction between the SFCs and the residential units to determine the allocation of storage spaces shared by the RUs. The auction price, on the other hand, is determined by a noncooperative Stackelberg game formulated between the residential units and the auctioneer. 

To design the scheme, as shown in Fig.~\ref{fig:ServiceDomainAuctionGame}, a smart community is considered that consists of a large number of residential units, which can be an individual home, or a large number of homes connected via an aggregator, and a number of SFCs that provide energy services such as managing lifts, corridor lights, water pumps, and heat pumps of the common facilities of the community. Each SFC and residential units have its own energy production capacity and storage devices. As the P2P is designed in \cite{Tushar_TSG_2016_sharing}, each SFC, which has larger energy generation capacity, may sometimes need larger storage space to store extra generation, and it can share storage spaces from the residential users of the community who has relatively small generation and storage capacity. Now, to facilitate this sharing (or, leasing) of storage spaces between multiple SFCs and residential units, in the designed modified auction process consist of three rules including a determination rule, a payment rule, and an allocation rule.

The objective of the determination rule is to determine the set of SFCs and residential units that can effectively participate in the auction scheme to determine the payment and shared storage amount. This is executed in a step by step fashion, and the number of participating SFC and residential units is affected by their respective bidding prices, a number of storage spaces that the SFCs want to share and the residential units agree to lease respectively, and the Vickrey price. Once the number of participating entities are determined, payment rule is executed to determine the auction price.

In payment rule, the proposed technique in \cite{Tushar_TSG_2016_sharing} is mainly varied from the Vickrey auction, and thus named as the modified auction scheme. In Vickrey auction, the auction price for sharing the storage spaces would be the second highest reservation price, i.e., the Vickrey price. However, this second highest price might not be considered beneficial by all the residential units participating in the auction scheme. Therefore, the auction price needs to be increased. On the other hand, if the auction price is set to the maximum bidding price, the price could be detrimental from some of the participating SFCs. Now, to make the auction scheme attractive and beneficial to all the participating residential units and, at the same time, to be cost effective for all the SFCs, a Stackelberg game between the auctioneer, which decides on the auction price to maximize the average cost savings to the SFCs as well as satisfying their desirable needs of storage spaces, and the residential units. Residential units decide on the vector containing the storage space they would like to put into the market for sharing such that their benefits are maximized. In the Stackelberg game, it is shown in that \emph{there always exists a unique solution to the game. Therefore, a unique auction price can be derived all the time, which all residential units and SFCs agree upon to be the equilibrium price for energy storage sharing between them.} Also, at this auction price, no participants have an incentive to deviate from the auction process.

Once the auction price is established, the allocation of storage spaces from the residential units to the SFCs is conducted based on an allocation rule. According to this rule, if the total requirement of the SFCs is either greater than or equal to the total supply from the residential units, then all the offered storage spaces are shared by the SFCs. However, if the supply is greater than the requirement, the participating residential units need to tolerate the burden of oversupply, i.e., the monetary loss in cases when the supply of storage becomes larger than the total requirement of storage spaces by the SFCs. In \cite{Tushar_TSG_2016_sharing}, two allocation processes are considered for the distribution of this burden: 1) proportional allocation and 2) equal allocation.
\begin{itemize}
\item\emph{Proportional allocation:} In proportional allocation, the burden of oversupply is shared among the residential units based on their respective reservation prices during the auction process. That is a residential unit, which asked for more reservation price will endure more burden compared to another residential unit with lower reservation price.
\item\emph{Equal allocation:} In equal allocation, however, the burden is distributed equally among all participating residential units.
\end{itemize}
The auction process is completed with the completion of the allocation process.

It is important to note that once an auction process is executed, there is always a possibility that the owners of the storage spaces, i.e., residential units, might cheat on the amount of storage that they agreed to put into the market during the auction. However, such auction schemes that possess the property of \emph{incentive compatibility} are secured from such cheating. Essentially, an auction process, in which no participant has any motivation to cheat during auction refers to as incentive compatible auction. In such an incentive compatible auction, the participants are satisfied with the allocation and payment that they received, a property known as \emph{individual rationality}, and therefore they have no incentive to cheat.  It is shown in \cite{Tushar_TSG_2016_sharing} that the Stackelberg game based payment rule and proportional allocation rule prepare the proposed modified auction for the P2P trading as individually rational. The scheme is further extended to a time-varying case, which also possesses all the properties of the static case as well.

Now, based on the above discussion, the overall exploration of modified auction scheme in the proposed storage sharing in the P2P trading network can be summarized as follows:
\begin{itemize}
\item[]\emph{\textbf{Step 1:}} The residential units and SFCs that can participate in the proposed auction are identified by following the determination rule.
\item[]\emph{\textbf{Step 2:}} The auction price is determined based on a Stackelberg game based payment rule. In the payment rule, it is shown that the derived auction price is unique, and all residential units and SFCs agree on that auction price for sharing energy storage spaces between them.
\item[]\emph{\textbf{Step 3:}} The allocation of storage space between the SFCs are conducted based on the allocation rule. The burden of oversupply, however, is distributed among the participating residential units using either an equal allocation or a proportional allocation scheme.
\item[]\emph{\textbf{Step 4:}} The proposed auction scheme is shown to be incentive compatible, and hence no participant has any incentive to cheat during the auction process. The property also holds when the auction scheme is extended to a time-varying case.
\end{itemize}

From the discussion in Section~\ref{sec:state-of-the-art} and Section~\ref{sec:basic-concepts}, the difference between the game-theoretic applications in existing energy management studies and in P2P energy network is obvious. In existing energy management, the focus of the studies is not primarily focusing on energy trading between them, rather cooperation or competition with one another to achieve an objective, which also involves the main grid significantly. On the contrary, in P2P, the participants also work together to achieve the desired objective, but with minimal (or no) interaction with the grid. A summary of these distinctive properties in this domain is shown in Table~\ref{table:TableSummery}.

\section{Outcomes of Game Theoretic Applications in P2P Energy Network}\label{sec:outcomes}
In Section~\ref{sec:basic-concepts}, we have provided detail descriptions of very specific game theoretic applications in P2P energy trading by discussing three different studies in detain in EV, DER and storage, and service domains respectively. In this section, our purpose is to study some interesting results from those studies, and show how the P2P scheme outperforms some existing schemes in these domains. These results, on the one hand, demonstrate the importance of P2P energy trading scheme in achieving greater benefit in cost reduction and utility maximization of energy entities within the network. On the other hand, these results also show the effectiveness of using game theory that enables the proposed scheme in achieving those benefits.
\subsection{EV Domain} 
\begin{figure*}[t]
\centering
\captionsetup{justification=centering}
\begin{minipage}[l]{0.5\textwidth}
\centering
\includegraphics[scale=0.4]{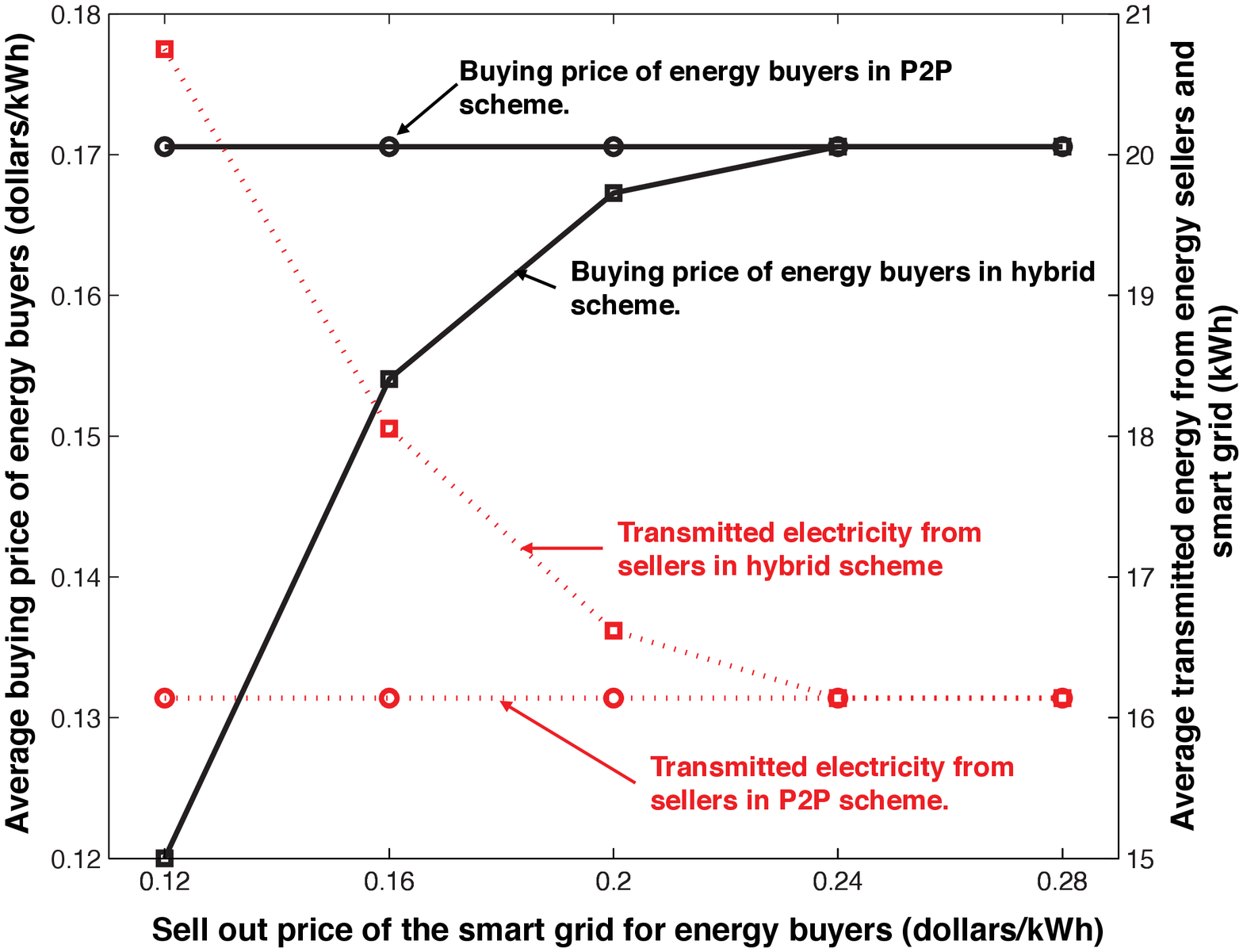}
\subcaption{\footnotesize{Average buying price and transmitted electricity for the EVs.}}
\label{fig:EVDomain_Fig1}
\end{minipage}
~
\begin{minipage}[r]{0.5\textwidth}
\centering
\captionsetup{justification=centering}
\includegraphics[scale=0.4]{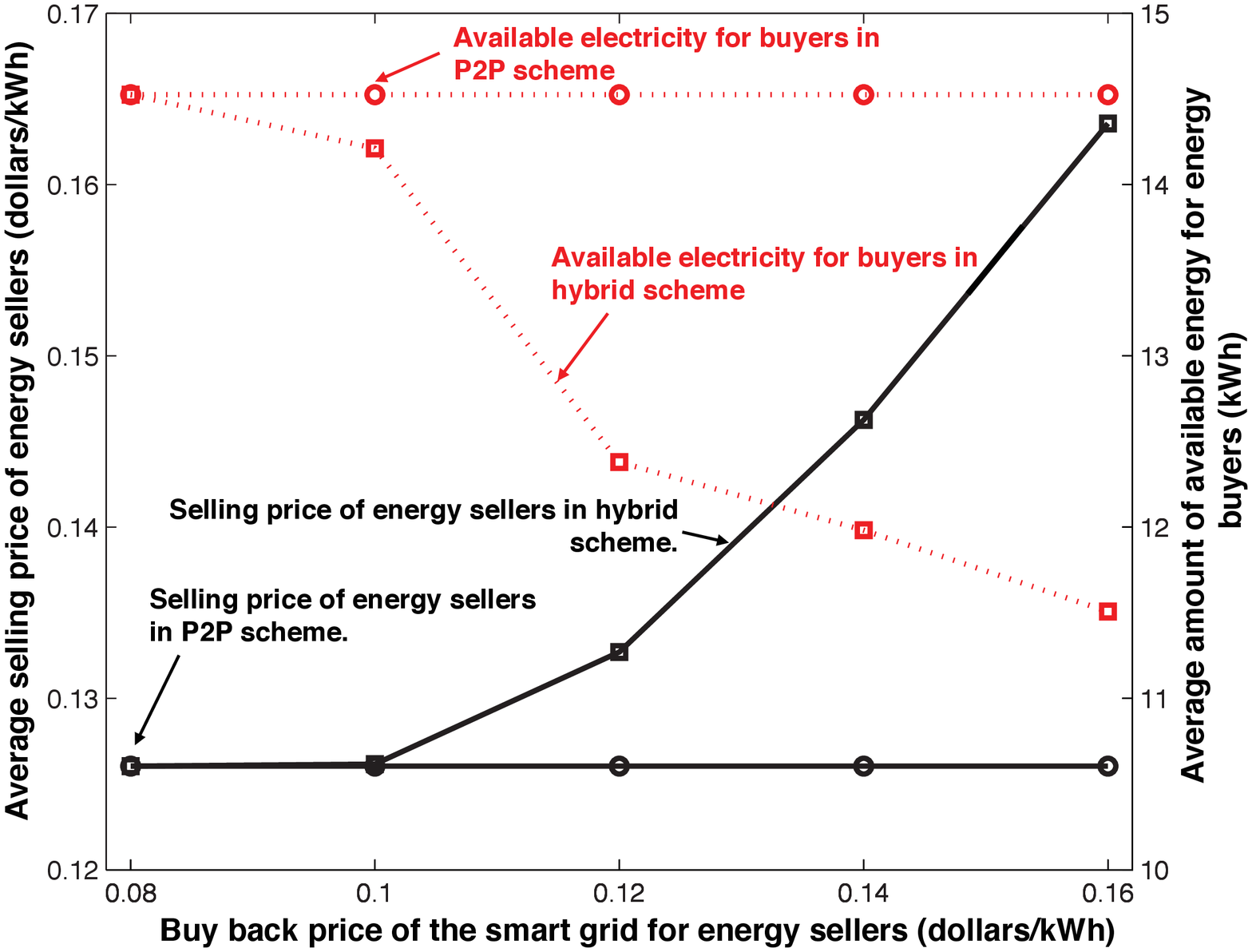}
\subcaption{\footnotesize{Average selling price and transmitted electricity for the EVs.}}
\label{fig:EVDomain_Fig2}
\end{minipage}
\captionsetup{justification=centering}
\caption{Both figures in this illustration are taken from \cite{Kang_TII_2017}, which provide insights on how the benefits of energy trading in P2P and hybrid models are influenced in terms of pricing and energy utilization.}
\label{fig:ResultP2PEV}
\end{figure*}

To demonstrate how game theory can be used to design P2P energy trading in EV domain, we have discussed an auction game based on \cite{Kang_TII_2017} in the previous section. Now, we show how the discussed game theoretic approach is beneficial for both the buyers and sellers of energy within the considered P2P network. To do so, we demonstrate two results from \cite{Kang_TII_2017} in Fig.~\ref{fig:ResultP2PEV}. Note that these performances are evaluated based on a real dataset in a real urban area of Texas. The latitude of observed area is from 30.256 to 30.276, and the longitude is from $-97.76$ to $-97.725$. The observed area is approximately $2.22 \times 3.88$ km$^2$ including $58$ parking lots. Therefore, the battery capacity of the EVs is set to 24 KWh, and the minimum and maximum of electricity demand for charging EVs are assumed to be [5, 10] KWh and [12, 18] KWh, respectively. The maximum of electricity supply for discharging EVs is considered to be [10, 20] KWh.

Fig. \ref{fig:ResultP2PEV} shows performance comparison between the proposed P2P model in \cite{Kang_TII_2017} and a hybrid energy trading model. By the hybrid energy trading model, the authors refer to the model in \cite{Wu_TSG_2015}, in which energy buyers can not only trade electricity with local energy seller but also with the smart grid. In \cite{Kang_TII_2017}, unlike the hybrid model, the focus is on localized P2P electricity trading between charging EVs (i.e., energy buyers) and discharging EVs (i.e., energy sellers) with 90\% electricity transmission efficiency in contrast to the high energy transmission losses between the smart grid and energy buyers and sellers in hybrid model resulting in low transmission efficiency of 70\%~ \cite{Kang_TII_2017}. Now, in Fig.~\ref{fig:EVDomain_Fig1}, it is shown that when the sell-out price of the smart grid for energy buyers is smaller than that of local discharging EVs, the energy buyers obtain more benefits by following a hybrid energy trading model because of the lower average buying price. However, because of high transmission losses, the average amount of transmitted electricity from both energy sellers and the smart grid is higher than that of the P2P scheme to meet the same requirement. If the sell-out price of the smart grid is too high, the energy buyers will buy electricity from local energy sellers instead of the smart grid in the hybrid model. Thus, they obtain the same benefits as the P2P scheme.

Similar results can also be found in Fig.~\ref{fig:EVDomain_Fig2}. Although the average selling price of energy sellers increases with the increasing buy-back price given by the smart grids, the average available electricity for energy buyers is decreasing because of higher energy losses during electricity transmission. Therefore, compared with the trading model in \cite{Wu_TSG_2015}, the proposed P2P model in \cite{Kang_TII_2017} has less energy loss and higher electricity utilization efficiency from the system's perspective. Thus, based on the results in Fig.~\ref{fig:ResultP2PEV}, it is reasonable to state that
\begin{itemize}
\item For both cases in Fig.~\ref{fig:EVDomain_Fig1} and \ref{fig:EVDomain_Fig2}, P2P energy trading is beneficial in terms of increasing systems' energy efficiency, which is mainly due to the lower transmission loss compared to the hybrid network.
\item In Fig.~\ref{fig:EVDomain_Fig1}, EVs would only be interested to participate in P2P trading when the sell-out price of the smart grid is very high, which is due to the fact that buyers are always motivated to buy from a source of energy that offers a lower price per unit of energy~\cite{Tushar-TIE:2014}. Therefore, to effectively establish such P2P trading scheme in the EV domain, the average buying price, which is $0.17$ dollar/kWh on average in the P2P network needs to be revised to a lower value to compete with the hybrid market. Nonetheless, the proposed P2P scheme is still able to attract EVs at peak period time when the electricity price is, in general, very high.
\item Due to a lower selling price on average, EVs may sell within P2P network during times when the price of hybrid network is also low. However, as the selling price increases in the hybrid network, more EVs would become interested to sell to the smart grid instead of in the P2P network. Hence, similar to the case in Fig.~\ref{fig:EVDomain_Fig1}, the selling price in P2P network should be chosen carefully. One example of a suitable pricing scheme for such P2P network could be mid-rate pricing scheme~\cite{Long_Conf_2017}.
\end{itemize}
\subsection{DER and Storage Domain}
\begin{figure*}[t]
\centering
\captionsetup{justification=centering}
\begin{minipage}[l]{0.5\textwidth}
\centering
\includegraphics[scale=0.33]{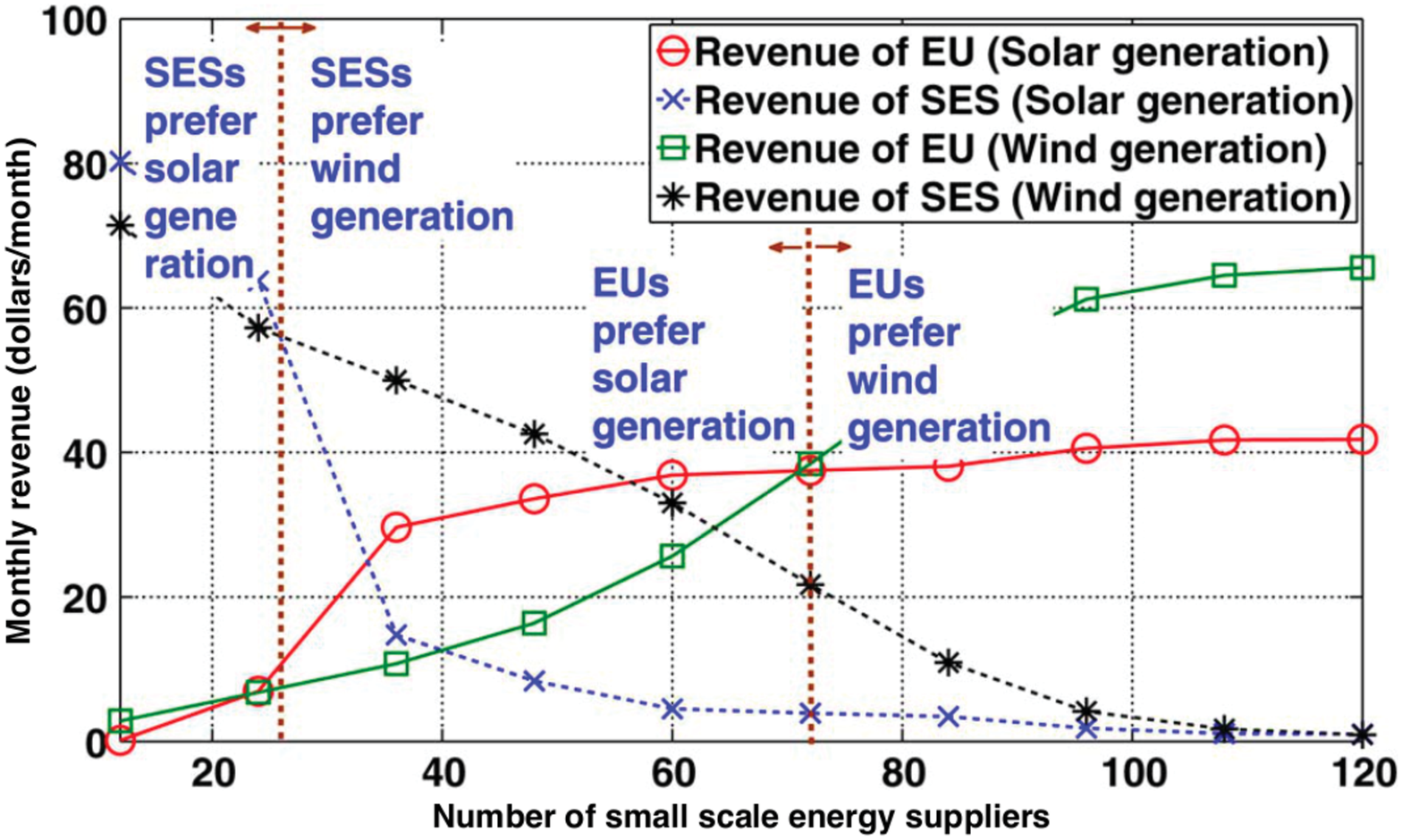}
\subcaption{\footnotesize{Effect of number of small scale energy suppliers on the monthly revenue of various end-users and energy suppliers.}}
\label{fig:DERDomain_Fig1}
\end{minipage}
~
\begin{minipage}[r]{0.5\textwidth}
\centering
\captionsetup{justification=centering}
\includegraphics[scale=0.35]{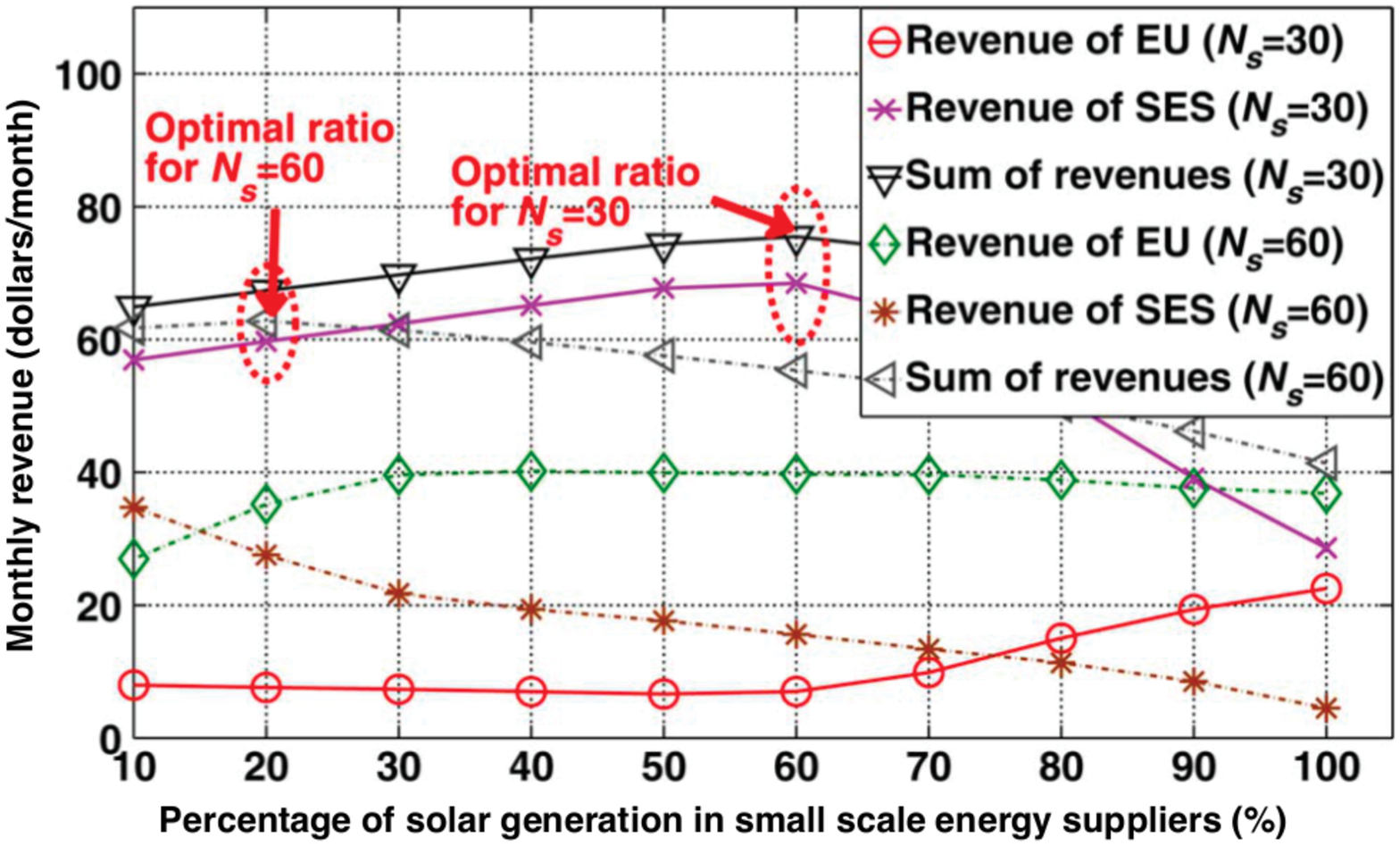}
\subcaption{\footnotesize{Effect of number of percentage of solar generation on the monthly revenue of various end-users and energy suppliers.}}
\label{fig:DERDomain_Fig2}
\end{minipage}
\captionsetup{justification=centering}
\caption{Both figures in this illustration are taken from \cite{Lee_JSAC_2014} as snapshots, which provide insights on how P2P energy trading may bring benefits to both end-users and small scale energy suppliers in terms of monetary revenue per month.}
\label{fig:ResultP2PDER}
\end{figure*}
In this section, we demonstrate how the P2P energy trading scheme can be beneficial in terms of earning revenue for both the buyers and sellers within the network. In particular, we discuss some results from \cite{Lee_JSAC_2014}, in which the load profile of households is constructed from the individual load profiles of home appliances. Each appliance has a different power consumption and a different probability to be activated in each hour of the day such that the load profile has different statistical characteristics (e.g., mean and variance) for different hours. The authors use an appliance load profile, which considers various appliances such as stove, dishwasher, refrigerator, and lighting. Then, they scale the load profile such that the average daily electricity usage of households is similar to that of households in North Carolina. 

As for the generation profile of small-scale energy supplies, solar generation and wind generation data profiles are used. For electricity generated by solar generation, the authors use the hourly electricity generation data which was measured at Elizabeth City State University in North Carolina during June 2012. The dataset was obtained from the Cooperative Networks For Renewable Resource Measurements website of the National Renewable Energy Laboratory (NREL). The generation profile of the solar generators is then scaled by assuming that the size of the solar panels is $6.45$ m$^2$. For electricity generated by wind turbines, the Eastern Wind Sources data set is used, which is also available at the NREL. The generation profile of the wind turbines is scaled by assuming that the capacity of the wind turbines is 5 kW.

Now, we demonstrate two outcomes from \cite{Lee_JSAC_2014} in Fig.~\ref{fig:ResultP2PDER}.  Fig.~\ref{fig:DERDomain_Fig1} shows the monthly revenue of individual end-users and small-scale energy suppliers who participate in P2P energy trading. In this figure, it is assumed that all energy suppliers have either solar generators or wind turbines. As shown in Fig.~\ref{fig:DERDomain_Fig1}, the monthly revenue of energy suppliers and end-users participating in P2P can reach up to $80$ and $62$ dollar, respectively. However, as shown in \cite{Lee_JSAC_2014}, the monthly electricity bill of a household without P2P reaches $110$ dollars, and therefore each end user can save up to $60\%$ of its monthly electricity bill by participating in P2P energy trading with one another.  Another phenomenon that we observe in Fig.~\ref{fig:DERDomain_Fig1} is that the monthly revenue of an energy supplier eventually decreases with increasing number of suppliers. This is mainly due to the characteristics of the P2P market as explained in \cite{Mengelkamp_AE_2017}. That is, as the number of small-scale energy suppliers increases in the market, the amount of available energy for sale increase subsequently, which leads to a drop in trading price. Therefore, the revenue to the suppliers decreases.

Further, Fig.~\ref{fig:DERDomain_Fig1} also shows how different kind of generation may affect the monthly revenue of the suppliers and end-users. Such scenario refers to the case in which the participants are able to choose the type of generation they would like to use for trading. Example of such a case can be found in the Brooklyn microgrid. Now, we observe in Fig.~\ref{fig:DERDomain_Fig1} that when the number of suppliers is lower than 24, both buyers and sellers prefer solar generation over wind generation due to higher monthly revenue of solar generation. Similarly, both parties prefer wind generation when the number of energy suppliers is larger than 72. However, if the number of energy suppliers is between 24 and 72, end-users prefer solar generation whereas energy suppliers prefer wind generation. Therefore, in this case, mixed usage of wind generators and solar generators is advisable.

Now, to get the answer to the question \emph{what is the suitable mix of solar and wind generations that would provide maximum benefit to the end-users and the small-scale energy suppliers}, Fig.~\ref{fig:DERDomain_Fig2} demonstrates the monthly monetary profit of end-users and energy suppliers for different percentages of solar generators used by energy suppliers.  According to this figure, there exists an optimal percentage of solar generators which maximizes the total revenue of end-users and energy suppliers. For example, when the number of suppliers is 30, the total revenue is maximized when 60\% of all suppliers are solar generators. It can also be seen from Fig.~\ref{fig:DERDomain_Fig2} that the optimal percentage of solar generators approaches zero as the number of energy suppliers increases, which is supported by the findings in Fig.~\ref{fig:DERDomain_Fig1}.

Based on the outputs in the DER and storage domain, as discussed above, we can summarize our insights as follows:
\begin{itemize}
\item Cooperation of participants within a P2P energy network is always beneficial. Because, it provides a platform to trade energy between themselves without involving the main grid, whose pricing scheme is not as attractive as the P2P scheme (as in the case of \cite{Lee_JSAC_2014}). Nonetheless, this may also be affected by how the pricing scheme is designed, as we discussed in the EV domain.
\item When the number of energy suppliers in the market becomes very large, the revenue to the energy suppliers reduces, which subsequently increases the revenue of the end-users of the P2P energy trading network.
\item Depending on the number of energy suppliers within the market, the different percentage mixture of solar and wind generations would be optimal for the energy suppliers to maximize their revenue from the energy trading.
\end{itemize}

\subsection{Service Domain}
\begin{figure}[t]
\centering
\captionsetup{justification=centering}
\includegraphics[width=0.6\textwidth]{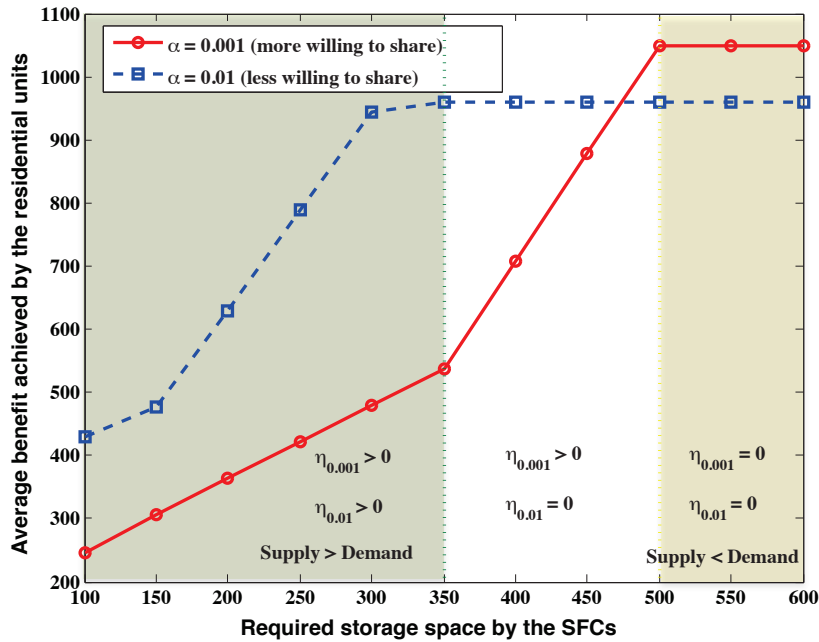}
\caption{Illustration of how the average benefit achieved by a residential unit may vary across various storage demands of the SFCs within the P2P energy network \cite{Tushar_TSG_2016_sharing}.}
\label{fig:ServiceDomain_Fig1}
\end{figure}
Finally, in this section, we will illustrate and discuss some of the findings of a game theoretic approach in the service domain, based on the study in \cite{Tushar_TSG_2016_sharing}. In this study, the authors consider a number of residential units at different blocks in a community that is interested in allowing the SFCs of the community to jointly share their energy storage devices and thus provide demand response services in the P2P energy trading market. When there are a large number of residential units and SFCs in the system, the reservation and bidding prices will vary significantly from one to another. Each residential unit is assumed to be a group of $[5, 25]$ households, where each household is equipped with a storage device of capacity $25$ kWh. The required electricity storage for each SFC is assumed to be within the range of $[100, 500]$ kWh. Nevertheless, the required storage space for sharing could be different if the users' usage pattern changes. Since the type of energy storage (and their associated cost) used by different residential units can vary significantly, the choices of reservation prices to share their storage space with the SFCs can vary considerably as well.

Note that once all the participating residential units put their free storage space that they would like to share into the auction market, they are distributed according to the allocation rule described in Section~\ref{sec:P2PServiceDomain}. In this regard, Fig.~\ref{fig:ServiceDomain_Fig1} investigate how the average utility of each residential unit is varied as the total storage amount required by the SFCs changes. In this case, the considered total ES requirement of the SFCs is assumed to be $100,150, 200, 250, 300, 350, 400, 450, 500, 550$ and $600$. Now, as shown in Fig. \ref{fig:ServiceDomain_Fig1}, in general, the average utility of each residential initially increases with increasing requirements of the SFCs and eventually becomes saturated to a stable value. This is due to the fact that as the required amount of storage space increases, the residential unit can share more of its reserved energy storage that it put into the market with the SFCs with the determined auction price. Hence, it's utility increases. However, each residential unit has a particular fixed storage amount that it can put on the market to share. Consequently, once the shared storage amount reaches its maximum, even with an increase in requirement of the SFCs the residential units cannot share more. Therefore, its utility becomes stable without any further increment. 

Another interesting observation from Fig.~\ref{fig:ServiceDomain_Fig1} is that the designed P2P storage sharing scheme favors the residential units with higher reluctance parameter more when the storage requirement of the SFCs is relatively lower and vice versa. Here, the reluctance parameter, which is denoted with $\alpha$ in Fig.~\ref{fig:ServiceDomain_Fig1}, refers to the measure of the willingness of sharing storage by each residential units. A lower value of $\alpha$ indicates higher willingness to share. Now, the variation of achieved benefit with different reluctance parameter is dictated by the burden of the oversupply of storage space. If reluctance is lower, a residential unit is interested to put a higher amount of storage into the market to share. However, if the total amount of energy storage required by the SFCs is lower, it would put a higher burden on the respective residential units.  As a consequence, the relative utility of the auction is lower. Nevertheless, if the requirement of the SFCs is higher, the sharing brings significant benefits to the residential units as can be seen from Fig. \ref{fig:ServiceDomain_Fig1}. On the other hand, with higher reluctance, residential units tend to share lower storage amount, which then enables them to endure a lower burden in case of lower demands from the SFCs. This consequently enhances their achieved utility. Nonetheless, if the requirement is higher from the SFCs, their utility reduces subsequently compared to the residential units with lower reluctance parameters. 
\begin{table}[t]
\centering
\captionsetup{justification=centering}
\caption{Demonstration of the improvement of average benefits obtained by the residential units in a P2P sharing scheme in comparison with ED and FiT schemes \cite{Tushar_TSG_2016_sharing}.}
\includegraphics[width=\textwidth]{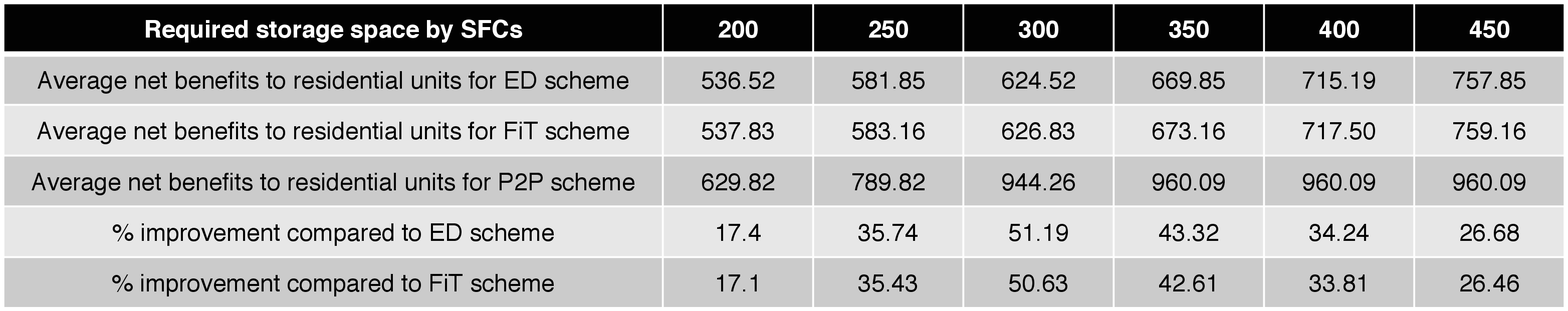}
\label{fig:ServiceDomain_Table1}
\end{table}

In Table~\ref{fig:ServiceDomain_Table1}, the resulting average utilities that each residential unit can achieve from sharing its storage space with the SFCs by adopting the P2P trading is shown and compared with existing equal distribution (ED), and feed-in-Tariff (FiT) schemes.

In the table, first, we note that as the amount of energy storage required by the SFCs increases, the average utility achieved per residential units also increases for all the cases due to the same reason explained in Fig.~\ref{fig:ServiceDomain_Fig1}. Also, in all the studied cases, the P2P storage sharing scheme shows a considerable performance improvement compared to both ED and FiT schemes. Particularly, an interesting trend of performance improvement can be observed if we compare the performance of the proposed scheme with the ED and FiT performances for each of the energy storage requirements. For instance, the performance of the P2P scheme is higher as the requirement for the storage increases from 200 to 350. However, the improvement is relatively less significant as the storage requirement switches from 400 to 450. This is mainly due to the fact that the amount of storage shared by each participating residential unit is influenced by their reluctance parameters. That is, even if the demand of the SFCs could be larger, the residential units may choose not to share more of their storage spaces if their reluctance is limited. Now, the residential units in the considered study increase their shares of energy storage as the requirement for the SFCs increases, which in turn produces higher revenue for them. Nonetheless, once the available storage spaces from the residential units reach the saturation, the increase in demand, i.e., from 400 to 450 in this case, does not affect their shares. As a consequence, their performance improvement is not as noticeable as in the previous four cases. Nonetheless, for all the considered cases, the auction process performs better than the ED scheme with an average performance improvement of 34.76\%, which clearly shows the value of the proposed methodology. The performance improvement with respect to the FiT scheme, which is 34.34\% on average, is due to the difference between the determined auction price and the price per unit of energy for the FiT scheme.

To this end, based on the above results, the key insights can be summarized as follows:
\begin{itemize}
\item From Fig.~\ref{fig:ServiceDomain_Fig1}, if the total required energy storage of the SFCs is smaller, residential units with higher reluctance benefit more and vice versa. This illustrates the fact that even residential units with high unwillingness to share their storage space can be beneficial for SFCs of the system if their required storage is small. 
\item For a higher storage requirement, SFCs would attain more benefit from having residential units with lower reluctances as they will be interested in sharing more to achieve higher average utilities.
\item Energy storage sharing in the P2P trading market is more beneficial for the residential units compared to the sharing by following both ED and FiT schemes.
\end{itemize}
\section{Conclusion}\label{sec:conclusion}In this paper, we have provided an overview of the potential of game theoretic approaches for energy management in the P2P network. To do so, first, we have highlighted the extensive use of game theoretic approaches in the smart energy domain and divided the discussion into three domains including EV domain, DER and storage domain and service domain.  Then, we have extended our focus on some recent game theoretic energy management models that have been proposed and implemented in P2P energy network. Here, instead of providing an overview, we have given a detail discussion of a specific game theoretic approach in each of the domain of the P2P network. The purpose has been to introduce the audience how different game theoretic models can be designed to solve energy trading problems in the P2P energy network, and what are the key criteria or properties that need to be considered during the implementation. Finally, we have shown some interesting results from the discussed game theoretic models and summarized the interpretation of those outcomes for a better understanding of participants' behavior in P2P energy management. 

The research of energy management in the P2P network is relatively new, and currently, all developments of P2P energy trading platforms are in pilot phase. Hence, a lot are yet to be done before integrating the P2P energy trading into the current energy system. In this context, some future research directions, in which game theory may play a significant role are provided as follows:
\begin{itemize}
\item\emph{Consumer-centric model:} The design of the P2P energy trading scheme needs to be consumer-centric. That is, consumers need to have benefit from participating in P2P energy trading. Note that some recent energy trading models and pilot projects have been discontinued as they were not accepted by the consumers. Hence, to avoid the occurrence of the same with P2P energy trading, the users' interests and benefits must be taken into consideration. One potential way to do this is to explore cooperative games to show that users can always benefit from cooperating one another. That is, a user may choose to be a part of the entire network (i.e., the grand coalition in a canonical coalition game) or dynamically change its position to a different small coalition (coalition formation game) to come to an agreement with other peers within the network for energy trading.
\item\emph{Demonstrated benefit to the grid:} In the most P2P energy trading pilots at present, the physical transfer of energy takes place over the distribution network which is set up by the traditional grid~\cite{Mengelkamp_AE_2017}. Hence, expecting that P2P energy trading will completely exclude the grid from any energy-related activities with the local consumers could be impractical, while the trading itself is conducted using the grid's asset. One potential way to address this problem is to demonstrate that P2P energy trading is also beneficial for the grid, and a grid may also participate in P2P energy trading, if necessary. This will also potentially help the regulatory board to understand the importance of P2P energy trading to both the grid and the local users, and thus pave the way for being approved for being a part of the energy system. The Stackelberg game could be an ideal candidate to model this trading, in which the grid can participate either as a leader or a follower, depending on the context of the model, and interact with other users to decide on various energy trading parameters across times.
\item\emph{High security and low computational complexity:} Due to the reduced involvement of centralized authority in P2P trading, the security, and privacy of participants has become a critical issue. In P2P network, an end user (buyer) does not want to reveal his/her identity during a transaction with a seller, whereas the seller does not want the buyer to misuse the traded energy, e.g., for illegal purposes. Therefore, there is a strong need for a energy trading distribution mechanism over P2P networks that do not pose security and privacy threats to the sellers and end users, respectively. The advancement of blockchain technology, however, has solved this problem. A blockchain is essentially a continuously growing list of records, called blocks, which are linked and secured using cryptography. The most existing pilot projects on P2P energy trading in the USA, Europe, and Australia are based on blockchain based information platform. Hence, how to integrate blockchain with game theory is a potential future research direction of significant importance. However, blockchain for privacy protection in peer-to-pee trading may require high computational power. Hence, the integration of blockchain with game theory needs to consider this characteristic with care, and design trading mechanisms that are efficient and possess less computational power to provide the desired service to the users.
\item\emph{Energy trading with incomplete information:} Incomplete information can be defined as  the lack of information of the real-time demand of prosumers and P2P trading price due to a problem in the network, e.g., packet loss in the communication network. Such incomplete information can potentially damage the performance of the P2P energy trading technique. Hence, there is a need to design energy management solutions that can properly deal with such scenarios. One promising  way to design energy trading mechanism for P2P network with incomplete information game. One example of such game is the Bayesian game whose solution is a Bayesian Nash Equilibrium.
\item\emph{Incorporation of physical laws in the game model:} An important aspect that governs the power flows on the network and couples DERs and aggregators on the physical network is the is the Kirchhoff laws, which are not properly modeled in most of the papers. It is important to note that the presence of the physical laws may greatly complicate the energy trading analysis and also has significant impact on how the market should be designed and operated. Hence, how to incorporate the impact of the Kirchhoff laws into the game theoretic model for peer-to-peer energy trading needs considerable attention. One potential way could be to include a common constraint between the players of the game, e.g., as usually considered in generalized Nash game, that will be influenced by the Kirchhoff laws. Nonetheless, in depth investigation is required to decide on how to introduce such a common coupling constraint.
\end{itemize} 
The potential application of game theoretic approaches in P2P energy trading and their subsequent implication on the participating users is large. The purpose of this paper has been to put a small drop in that large vessel by showing the importance of game theory for such a network via demonstrating what game theory is capable of and how it has been used so far, and to provide the reader with some fruits for thoughts on how they might contribute in this emerging energy domain by using game theory.
\section*{Acknowledgement}This work is supported in part by the Advance Queensland Research Fellowship AQRF11016-17RD2, which is jointly sponsored by the State of Queensland through the Department of Science, Information Technology and Innovation, the University of Queensland and Redback Technologies; in part by the project NRF2015ENC-GBICRD001-028 funded by National Research Foundation (NRF) via the Green Buildings Innovation Cluster (GBIC), which is administered by Building and Construction Authority (BCA); in part by the SUTD-MIT International Design Centre (idc; idc.sutd.edu.sg); and in part by the NSF grant 1253516.

\begin{thebibliography}{10}
\providecommand{\url}[1]{#1}
\csname url@samestyle\endcsname
\providecommand{\newblock}{\relax}
\providecommand{\bibinfo}[2]{#2}
\providecommand{\BIBentrySTDinterwordspacing}{\spaceskip=0pt\relax}
\providecommand{\BIBentryALTinterwordstretchfactor}{4}
\providecommand{\BIBentryALTinterwordspacing}{\spaceskip=\fontdimen2\font plus
\BIBentryALTinterwordstretchfactor\fontdimen3\font minus
  \fontdimen4\font\relax}
\providecommand{\BIBforeignlanguage}[2]{{%
\expandafter\ifx\csname l@#1\endcsname\relax
\typeout{** WARNING: IEEEtran.bst: No hyphenation pattern has been}%
\typeout{** loaded for the language `#1'. Using the pattern for}%
\typeout{** the default language instead.}%
\else
\language=\csname l@#1\endcsname
\fi
#2}}
\providecommand{\BIBdecl}{\relax}
\BIBdecl

\bibitem{Tushar-TIE:2014}
W.~Tushar, B.~Chai, C.~Yuen, D.~B. Smith, K.~L. Wood, Z.~Yang, and H.~V. Poor,
  ``Three-party energy management with distributed energy resources in smart
  grid,'' \emph{IEEE Transactions on Industrial Electronics}, vol.~62, no.~4,
  pp. 2487--2498, Apr. 2015.

\bibitem{Gan_TPS_2013}
L.~Gan, U.~Topcu, and S.~H. Low, ``Optimal decentralized protocol for electric
  vehicle charging,'' \emph{IEEE Transactions on Power Systems}, vol.~28,
  no.~2, pp. 940--951, May 2013.

\bibitem{Mohsenian-Rad_TSG_2010}
A.~H. Mohsenian-Rad, V.~W.~S. Wong, J.~Jatskevich, R.~Schober, and
  A.~Leon-Garcia, ``Autonomous demand-side management based on game-theoretic
  energy consumption scheduling for the future smart grid,'' \emph{IEEE
  Transactions on Smart Grid}, vol.~1, no.~3, pp. 320--331, Dec. 2010.

\bibitem{Boss_IEEE_2017}
B.~K. Bose, ``Artificial intelligence techniques in smart grid and renewable
  energy systems - {Some} example applications,'' \emph{Proceedings of the
  IEEE}, vol. 105, no.~11, pp. 2262--2273, Nov. 2017.

\bibitem{Saad_GameSmartGrid_2012}
W.~Saad, Z.~Han, H.~V. Poor, and T.~Ba\c{s}ar, ``Game-theoretic methods for the
  smart grid: An overview of microgrid systems, demand-side management, and
  smart grid communications,'' \emph{IEEE Signal Processing Magazine}, vol.~29,
  no.~5, pp. 86--105, Sep. 2012.

\bibitem{Voyant_RE_2017}
C.~Voyant, G.~Notton, S.~Kalogirou, M.-L. Nivet, C.~Paoli, F.~Motte, and
  A.~Fouilloy, ``Machine learning methods for solar radiation forecasting: {A}
  review,'' \emph{Renewable Energy}, vol. 105, pp. 569--582, May 2017.

\bibitem{Peck_IEEE_Spectrum_2017}
M.~E. Peck and D.~Wagman, ``Energy trading for fun and profit buy your
  neighbor's rooftop solar power or sell your own-it'll all be on a
  blockchain,'' \emph{IEEE Spectrum}, vol.~54, no.~10, pp. 56--61, Oct. 2017.

\bibitem{FiTPolicy_2015}
{Strategic Futures, Energy Industry Policy}, ``{Queensland solar bonus scheme
  policy guide},'' Department of Energy and Water Supply, State of Queensland,
  QLD, Australia, Report, 2015.

\bibitem{Mengelkamp_AE_2017}
E.~Mengelkamp, J.~G{\"a}rttner, K.~Rock, S.~Kessler, L.~Orsini, and
  C.~Weinhardt, ``Designing microgrid energy markets - {A} case study: {The
  Brooklyn} microgrid,'' \emph{Applied Energy}, pp. 1--11, June 2017,
  pre-print.

\bibitem{Economist_2013}
\BIBentryALTinterwordspacing
{The Economist}, ``{The rise of sharing economy},'' Mar. 2013. [Online].
  Available:
  \url{http://www.economist.com/news/leaders/21573104-internet-everything-hire-rise-sharing-economy}
\BIBentrySTDinterwordspacing

\bibitem{Wayes-J-TSG:2012}
W.~Tushar, W.~Saad, H.~V. Poor, and D.~B. Smith, ``Economics of electric
  vehicle charging: {A} game theoretic approach,'' \emph{IEEE Transactions on
  Smart Grid}, vol.~3, no.~4, pp. 1767--1778, Dec 2012.

\bibitem{Tushar_TSG_2017}
W.~Tushar, C.~Yuen, D.~B. Smith, and H.~V. Poor, ``Price discrimination for
  energy trading in smart grid: A game theoretic approach,'' \emph{IEEE
  Transactions on Smart Grid}, vol.~8, no.~4, pp. 1790--1801, July 2017.

\bibitem{Marzband_IET_2016}
M.~Marzband, M.~Javadi, J.~L. Dom{\'i}nguez-Garc{\'i}a, and M.~M. Moghaddam,
  ``Non-cooperative game theory based energy management systems for energy
  district in the retail market considering {DER} uncertainties,'' \emph{IET
  Generation, Transmission and Distribution}, vol.~10, no.~12, pp. 2999--3009,
  Sep. 2016.

\bibitem{Wu_TSG_2012}
C.~Wu, H.~Mohsenian-Rad, and J.~Huang, ``Vehicle-to-aggregator interaction
  game,'' \emph{IEEE Transactions on Smart Grid}, vol.~3, no.~1, pp. 434--442,
  Mar. 2012.

\bibitem{Nguyen_JCN_2012}
H.~K. Nguyen and J.~B. Song, ``Optimal charging and discharging for multiple
  {PHEVs} with demand side management in vehicle-to-building,'' \emph{Journal
  of Communications and Networks}, vol.~14, no.~6, pp. 662--671, Dec. 2012.

\bibitem{Bacci_SPM_2016}
G.~Bacci, S.~Lasaulce, W.~Saad, and L.~Sanguinetti, ``{Game theory for networks
  - A tutorial on game-theoretic tools for emerging signal processing
  applications},'' \emph{IEEE Signal Processing Magazine}, vol.~33, no.~1, pp.
  94--119, Jan. 2016.

\bibitem{Saad_CoopGame_2009}
W.~Saad, Z.~Han, M.~Debbah, A.~Hj{\o}rungnes, and T.~Ba\c{s}ar, ``Coalitional
  game theory for communication networks,'' \emph{IEEE Signal Processing
  Magazine}, vol.~26, no.~5, pp. 77--97, Sep. 2009.

\bibitem{Liu_TSG_2017}
Z.~Liu, Q.~Wu, S.~Huang, L.~Wang, M.~Shahidehpour, and Y.~Xue, ``Optimal
  day-ahead charging scheduling of electric vehicles through an aggregative
  game model,'' \emph{IEEE Transactions on Smart Grid}, vol.~PP, no.~99, pp.
  1--1, 2017.

\bibitem{Lee_TSG_2015}
W.~Lee, L.~Xiang, R.~Schober, and V.~W.~S. Wong, ``Electric vehicle charging
  stations with renewable power generators: {A} game theoretical analysis,''
  \emph{IEEE Transactions on Smart Grid}, vol.~6, no.~2, pp. 608--617, Mar.
  2015.

\bibitem{Zou_TAC_2017}
S.~Zou, Z.~Ma, X.~Liu, and I.~Hiskens, ``An efficient game for coordinating
  electric vehicle charging,'' \emph{IEEE Transactions on Automatic Control},
  vol.~62, no.~5, pp. 2374--2389, May 2017.

\bibitem{Wang_TSG_2014}
Y.~Wang, W.~Saad, Z.~Han, H.~V. Poor, and T.~Ba\c{s}ar, ``A game-theoretic
  approach to energy trading in the smart grid,'' \emph{IEEE Transactions on
  Smart Grid}, vol.~5, no.~3, pp. 1439--1450, May 2014.

\bibitem{Kang_TII_2017}
J.~Kang, R.~Yu, X.~Huang, S.~Maharjan, Y.~Zhang, and E.~Hossain, ``Enabling
  localized peer-to-peer electricity trading among plug-in hybrid electric
  vehicles using consortium blockchains,'' \emph{IEEE Transactions on
  Industrial Informatics}, pp. 1--10, 2017, pre-print.

\bibitem{Kumar_TDSC_2016}
N.~Kumar, S.~Misra, N.~Chilamkurti, J.~H. Lee, and J.~J. P.~C. Rodrigues,
  ``Bayesian coalition negotiation game as a utility for secure energy
  management in a vehicles-to-grid environment,'' \emph{IEEE Transactions on
  Dependable and Secure Computing}, vol.~13, no.~1, pp. 133--145, Jan. 2016.

\bibitem{Yu_ITJ_2014}
R.~Yu, J.~Ding, W.~Zhong, Y.~Liu, and S.~Xie, ``Phev charging and discharging
  cooperation in v2g networks: A coalition game approach,'' \emph{IEEE Internet
  of Things Journal}, vol.~1, no.~6, pp. 578--589, Dec. 2014.

\bibitem{Zhao_TSG_2017}
T.~Zhao, Y.~Li, X.~Pan, P.~Wang, and J.~Zhang, ``Real-time optimal energy and
  reserve management of electric vehicle fast charging station: Hierarchical
  game approach,'' \emph{IEEE Transactions on Smart Grid}, vol.~PP, no.~99, pp.
  1--1, 2017.

\bibitem{Wang_TVT_2017}
R.~Wang, G.~Xiao, and P.~Wang, ``Hybrid centralized-decentralized {(HCD)}
  charging control of electric vehicles,'' \emph{IEEE Transactions on Vehicular
  Technology}, vol.~66, no.~8, pp. 6728--6741, Aug. 2017.

\bibitem{Yuan_TSG_2017}
W.~Yuan, J.~Huang, and Y.~J.~A. Zhang, ``Competitive charging station pricing
  for plug-in electric vehicles,'' \emph{IEEE Transactions on Smart Grid},
  vol.~8, no.~2, pp. 627--639, Mar. 2017.

\bibitem{Yang1_TVT_2016}
H.~Yang, X.~Xie, and A.~V. Vasilakos, ``Noncooperative and cooperative
  optimization of electric vehicle charging under demand uncertainty: A robust
  stackelberg game,'' \emph{IEEE Transactions on Vehicular Technology},
  vol.~65, no.~3, pp. 1043--1058, Mar. 2016.

\bibitem{Tan_TSG_2017}
J.~Tan and L.~Wang, ``A game-theoretic framework for vehicle-to-grid frequency
  regulation considering smart charging mechanism,'' \emph{IEEE Transactions on
  Smart Grid}, vol.~8, no.~5, pp. 2358--2369, Sept. 2017.

\bibitem{Mondol_IET_2015}
A.~Mondal and S.~Misra, ``Game-theoretic energy trading network topology
  control for electric vehicles in mobile smart grid,'' \emph{IET Networks},
  vol.~4, no.~4, pp. 220--228, 2015.

\bibitem{Zhu_Access_2016}
Z.~Zhu, S.~Lambotharan, W.~H. Chin, and Z.~Fan, ``A mean field game theoretic
  approach to electric vehicles charging,'' \emph{IEEE Access}, vol.~4, pp.
  3501--3510, 2016.

\bibitem{Chen_TSG_2014}
H.~Chen, Y.~Li, R.~H.~Y. Louie, and B.~Vucetic, ``Autonomous demand side
  management based on energy consumption scheduling and instantaneous load
  billing: An aggregative game approach,'' \emph{IEEE Transactions on Smart
  Grid}, vol.~5, no.~4, pp. 1744--1754, July 2014.

\bibitem{Atzeni_TSG_2013}
I.~Atzeni, L.~G. Ord{\`o}{\~n}ez, G.~Scutari, D.~P. Palomar, and J.~R.
  Fonollosa, ``Demand-side management via distributed energy generation and
  storage optimization,'' \emph{IEEE Transactions on Smart Grid}, vol.~4,
  no.~2, pp. 866--876, June 2013.

\bibitem{Cintuglu_TSG_2015}
M.~H. Cintuglu, H.~Martin, and O.~A. Mohammed, ``Real-time implementation of
  multiagent-based game theory reverse auction model for microgrid market
  operation,'' \emph{IEEE Transactions on Smart Grid}, vol.~6, no.~2, pp.
  1064--1072, Mar. 2015.

\bibitem{Tushar_TSG_2016_sharing}
W.~Tushar, B.~Chai, C.~Yuen, S.~Huang, D.~B. Smith, H.~V. Poor, and Z.~Yang,
  ``Energy storage sharing in smart grid: A modified auction-based approach,''
  \emph{IEEE Transactions on Smart Grid}, vol.~7, no.~3, pp. 1462--1475, May
  2016.

\bibitem{Lee_JSAC_2014}
W.~Lee, L.~Xiang, R.~Schober, and V.~W.~S. Wong, ``Direct electricity trading
  in smart grid: {A} coalitional game analysis,'' \emph{IEEE Journal on
  Selected Areas in Communications}, vol.~32, no.~7, pp. 1398--1411, July 2014.

\bibitem{Ni_IET_2016}
J.~Ni and Q.~Ai, ``{Economic power transaction using coalitional game strategy
  in micro-grids},'' \emph{IET Generation, Transmission \& Distribution},
  vol.~10, no.~1, pp. 10--18, Jan. 2016.

\bibitem{Maharjan_TSG_2013}
S.~Maharjan, Q.~Zhu, Y.~Zhang, S.~Gjessing, and T.~Ba{\c{s}}ar, ``Dependable
  demand response management in the smart grid: {A Stackelberg} game
  approach,'' \emph{IEEE Transactions on Smart Grid}, vol.~4, no.~1, pp.
  120--132, Mar. 2013.

\bibitem{Nava_TSMCS_2014}
E.~Mojica-Nava, C.~A. Macana, and N.~Quijano, ``Dynamic population games for
  optimal dispatch on hierarchical microgrid control,'' \emph{IEEE Transactions
  on Systems, Man, and Cybernetics: Systems}, vol.~44, no.~3, pp. 306--317,
  Mar. 2014.

\bibitem{Cintuglu_TII_2017}
M.~H. Cintuglu and O.~A. Mohammed, ``Behavior modeling and auction architecture
  of networked microgrids for frequency support,'' \emph{IEEE Transactions on
  Industrial Informatics}, vol.~13, no.~4, pp. 1772--1782, Aug. 2017.

\bibitem{Zhu_Conf_2011}
Q.~Zhu and T.~Ba\c{s}ar, ``A multi-resolution large population game framework
  for smart grid demand response management,'' in \emph{Proc. of the
  International Conference on NETwork Games, Control and Optimization (NetGCooP
  2011)}, Paris, France, Oct. 2011, pp. 1--8.

\bibitem{Giotitsas_2015}
C.~Giotitsas, A.~Pazaitis, and V.~Kostakis, ``{A peer-to-peer approach to
  energy production},'' \emph{Technology in Society}, vol.~42, pp. 28--38, Aug.
  2015.

\bibitem{Llic_Conf_2012}
D.~Ilic, P.~G.~D. Silva, S.~Karnouskos, and M.~Griesemer, ``An energy market
  for trading electricity in smart grid neighbourhoods,'' in \emph{IEEE
  International Conference on Digital Ecosystems and Technologies (DEST)},
  Campione d'Italia, Italy, July 2012, pp. 1--6.

\bibitem{Saad_SmartGridComm_2011}
W.~Saad, Z.~Han, H.~V. Poor, and T.~Ba{\c{s}}ar, ``{A noncooperative game for
  double auction-based energy trading between PHEVs and distribution grids},''
  in \emph{Proc. of the IEEE International Conference on Smart Grid
  Communications (SmartGridComm)}, Brussels, Belgium, Oct. 2011, pp. 267--272.

\bibitem{Alvaro-Hermana_TSM_2016}
R.~Alvaro-Hermana, J.~Fraile-Ardanuy, P.~J. Zufiria, L.~Knapen, and
  D.~Janssens, ``Peer to peer energy trading with electric vehicles,''
  \emph{IEEE Intelligent Transportation Systems Magazine}, vol.~8, no.~3, pp.
  33--44, Fall 2016.

\bibitem{Samadi_TSG_2016}
P.~Samadi, V.~W.~S. Wong, and R.~Schober, ``Load scheduling and power trading
  in systems with high penetration of renewable energy resources,'' \emph{IEEE
  Transactions on Smart Grid}, vol.~7, no.~4, pp. 1802--1812, July 2016.

\bibitem{Kalathil_TSG_2017}
D.~Kalathil, C.~Wu, K.~Poolla, and P.~Varaiya, ``The sharing economy for the
  electricity storage,'' \emph{IEEE Transactions on Smart Grid}, vol.~PP,
  no.~99, pp. 1--1, 2017.

\bibitem{Zhang_EP_2016}
C.~Zhang, J.~Wu, M.~Cheng, Y.~Zhou, and C.~Long, ``A bidding system for
  peer-to-peer energy trading in a gridconnected microgrid,'' \emph{Energy
  Procedia}, vol. 103, pp. 147--152, Dec. 2016.

\bibitem{Wu_TSG_2015}
Y.~Wu, X.~Tan, L.~Qian, D.~H.~K. Tsang, W.~Z. Song, and L.~Yu, ``Optimal
  pricing and energy scheduling for hybrid energy trading market in future
  smart grid,'' \emph{IEEE Transactions on Industrial Informatics}, vol.~11,
  no.~6, pp. 1585--1596, Dec. 2015.

\bibitem{Long_Conf_2017}
C.~Long, J.~Wu, C.~Zhang, L.~Thomas, M.~Cheng, and N.~Jenkins, ``{Peer-to-peer
  energy trading in a community microgrid},'' in \emph{Proc. IEEE PES General
  Meeting}, Chicago, IL, July 2017, pp. 1--5.
\end{thebibliography}

\end{document}